\documentclass[aps,prb,twocolumn,longbibliography,groupedaddress,showpacs,floatfix,superscriptaddress]{revtex4-1}
\usepackage{graphicx}  
\usepackage{dcolumn}  
\usepackage[english]{babel}
\usepackage{bm}  
\usepackage{amsmath, amsthm, amssymb}
\usepackage{bbold}
\usepackage{mathrsfs}
\usepackage{array}
\usepackage[usenames]{color}
\usepackage{soul} 
\usepackage{ulem} 
\usepackage{float}

\usepackage{hyperref}
\hypersetup{
    colorlinks=true,
    linkcolor=blue,
    filecolor=magenta,      
    urlcolor=blue,
    pdftitle={Overleaf Example},
    pdfpagemode=FullScreen,
    }

\emergencystretch=\maxdimen
\newcommand{\calH}{\mathcal{H}}
\newcommand{\iv}{\mathbf{i}}
\newcommand{\jv}{\mathbf{j}}
\DeclareUnicodeCharacter{2212}{-}

\begin{document}

\normalem

\title{Magnetic and singlet phases in the three-dimensional periodic Anderson Model}
\author{Wiliam S. Oliveira}
\affiliation{Instituto de F\'isica, Universidade Federal do Rio de
Janeiro Cx.P. 68.528, 21941-972 Rio de Janeiro RJ, Brazil}
\author{Thereza Paiva} 
\affiliation{Instituto de F\'isica, Universidade Federal do Rio de
Janeiro Cx.P. 68.528, 21941-972 Rio de Janeiro RJ, Brazil}
\author{Richard T. Scalettar} 
\affiliation{Department of Physics and Astronomy, University of California, Davis, CA 95616 USA}
\author{Natanael C. Costa}
\affiliation{Instituto de F\'isica, Universidade Federal do Rio de
Janeiro Cx.P. 68.528, 21941-972 Rio de Janeiro RJ, Brazil}

\begin{abstract}
Heavy fermion materials are compounds in which localized $f$-orbitals hybridize with delocalized $d$ ones, leading to quasiparticles with large renormalized masses. The presence of strongly correlated $f$-electrons at the Fermi level may also lead to long-range order, such as magnetism, or unconventional superconductivity. From a theoretical point of view, the ``standard model'' for heavy fermion compounds is the Periodic Anderson Model (PAM).  Despite being extensively scrutinized, its thermodynamic properties in three-dimensional lattices have not been carefully addressed by unbiased methodologies. Here we investigate the 3D PAM employing state-of-the-art finite temperature auxiliary field quantum Monte Carlo simulations. We present the behavior of the kinetic energy, the entropy, the specific heat, and the double occupancy as functions of the temperature and the hybridization strength. From these quantities, and by the analysis of the spin-spin correlation functions, we investigate the occurrence of magnetic phase transitions at finite temperatures, and determine the phase diagram of the model, including the behavior of the N\'eel temperature as a function of the external parameters.
\end{abstract}

%\date{Version 1.00 -- \today}

\pacs{
71.10.Fd, % Lattice fermion models (Hubbard model, etc.)
02.70.Uu  % Applications of Monte Carlo methods
}
\maketitle

%\tcm{comments by TP}
%%%%%%%%%%%%%%%%%%%%%%%%%%%%%%%%%%%%%%%%%%%%%%%%%%%%%%%%%%%%%%%%%%
%%%%%%%%%%%%%%%%%%%%%%%%%%%%%%%%%%%%%%%%%%%%%%%%%%%%%%%%%%%%%%%%%%
\section{Introduction}
%%%%%%%%%%%%%%%%%%%%%%%%%%%%%%%%%%%%%%%%%%%%%%%%%%%%%%%%%%%%%%%%%%
%%%%%%%%%%%%%%%%%%%%%%%%%%%%%%%%%%%%%%%%%%%%%%%%%%%%%%%%%%%%%%%%%%

Real-space Quantum Monte Carlo (QMC) studies of itinerant electron and electron-phonon models, such as the 
Hubbard, Holstein, Su-Schrieffer-Heeger (SSH), and Periodic Anderson Hamiltonians, have often focused on one and two dimensions \cite{Raczkowski20,Paleari21,Xing21,Hu19}.
In the case of the Hubbard model, this choice is partly driven by direct implications in quasi-2D compounds, e.g.
%% , the 2D case is the one which is likely 
layered geometries relevant to cuprate materials and $d$-wave superconductivity\,\cite{Scalapino12}.  
Similarly, other 2D materials/lattices, such as graphene and transition metal dichalcogenides, may exhibit interesting topological properties or long-range ordered phases\,\cite{Kotov12,Manzeli17}.
However, the choice 
%% for 2D 
of lower dimensional lattices is also a practical one: with the system sizes limited to ${\cal O}(10^2$-$10^3)$ sites, reliable finite-size scaling analysis becomes challenging in 3D geometries.
Unfortunately, due to the Mermin-Wagner theorem, lower dimension may also lead to restrictions for the occurrence of long-range ordered phases.
This is the case of Hubbard-like models, which at half-filling any possible long-range antiferromagnetic phase should occur only at zero temperature.
%Unfortunately, lower dimension also means the reduction of the possibility of long-range order, since the continuous spin symmetry of the Hubbard model, for example, implies that at half-filling any possible long-range antiferromagnetic phase should occur only at zero temperature.
\footnote{The reduced (Ising) order parameter symmetry of the Holstein case does support a finite $T$ transition in 2D.}
Studies within dynamical mean-field theory (DMFT), dynamical cluster approximation, or diagrammatic Monte Carlo avoid this issue,
%% of low dimensionality, 
while introducing other approximations or limitations.

A significant exception is the 
%% Despite this, a 
great numerical effort 
%% has been done to investigate 
devoted to investigating the 3D (cubic) Hubbard\,\cite{hirsch87,scalettar89,Muramatsu00,fuchs11} model, the 3D (cubic) Holstein\,\cite{cohenstead20}, and (cubic perovskite) SSH models\,\cite{cohenstead22} within QMC simulations.
These studies have established quantitatively the dependence of their critical temperatures at half-filling -- $T_N$ (N\'eel) to the Hubbard model, and $T_{\rm cdw}$ (CDW) to the Holstein model.
Such a difficult task is alleviated by the increasing computational resources, with parallelization protocols, as well as the development of new methodologies (e.g., machine learning or Langevin accelerations).
Although this leap to 3D geometries is a significant accomplishment, so far it has been carried out mostly for single-band models.

As a paradigm for two-band systems, the periodic Anderson model (PAM) considers the interplay of localized and itinerant fermions.
% is a natural next target for 3D simulations.
Unlike the half-filled 3D Hubbard model, where one has ground state antiferromagnetic even for $U_{f}/t \to 0$, in the PAM there is a finite critical ratio of interband to conduction electron hybridization $V/t$, separating antiferromagnetic and singlet phases at $T=0$.
Thus, one has a substantially more rich phase diagram.
%% , which is unknown for 
However its quantitative critical boundaries have not been determined by
unbiased methodologies at the present moment.
Early attempts to describe the properties of the 3D PAM by QMC was performed in Refs.\onlinecite{McMahan98,Huscroft99,Paiva03}, in which the behavior of the spin-spin correlation functions was examined for different values of $V/t$ and $U_{f}/t$, but for small lattices and without a finite-size scaling analysis.  
%% Therefore, 
The main goal of this paper is to bridge this gap by extending work on real space QMC in 3D to the PAM, and comparing it with the well-known 2D case.

%Studies of the 2D and 3D PAM with nearest-neighbor hybridization \cite{Huscroft99} have shown a crossover from an antiferromagnetic to a singlet state for $V/t \simeq 0.6-0.8$. 
%We should also cite \cite{Schafer19}, dynamical vertex approximation for the 2D PAM, shows a quantum phase transition between and AF and a Kondo insulator at $V_c/t \approx 0.91$.

As the standard model for heavy-fermion compounds, the study of the 3D PAM is a significant step towards understanding the thermodynamic properties of these materials\,\cite{bernhard99}.
In particular, the application of pressure changes the conduction electron bandwidth $W=12\,t$ and hybridization $V$ between localized and conduction bands (with a lesser effect on on-site repulsion).
These changes affect the nature of both high and low-temperature responses, from the transition temperatures to transport properties.
Identifying these energy scales is crucial since, given a density-functional theory parameterization of $W$ and $V$, our phase transitions can be used for quantitative benchmarking of heavy fermion properties -- such as the Cerium volume collapse transition\,\cite{held01}.
The present paper is organized as follows:
In Sec.~\ref{sec:HQMC} we present the Periodic Anderson model, and highlight the main features of the QMC method together with the quantities of interest.
Our results are presented in Sec.~\ref{sec:results}, while the main conclusions are summarized in Sec.~\ref{sec:conc}.

%%%%%%%%%%%%%%%%%%%%%%%%%%%%%%%%%%%%%%%%%%%%%%%%%%%%%%%%%%%%%%%%%%
\section{Model and Method}
\label{sec:HQMC}
%%%%%%%%%%%%%%%%%%%%%%%%%%%%%%%%%%%%%%%%%%%%%%%%%%%%%%%%%%%%%%%%%%
The PAM Hamiltonian reads

\begin{align}\label{eq:hamil}
\hat	\calH=&-t\sum_{\langle\iv,\jv\rangle,\sigma}\left(d_{\iv\sigma}^\dagger 
d_{\jv\sigma}^{\phantom{\dagger}}+\mathrm{h.c.}\right)
	-V\sum_{\iv,\sigma}\left(d_{\iv\sigma}^\dagger 
f_{\iv\sigma}^{\phantom{\dagger}}+\mathrm{H.c.}\right)\nonumber\\
&-\mu \sum_{\iv,\sigma,\alpha} n^{\alpha}_{\iv\sigma}
+U_f\sum_{\iv}
\left(n^{f}_{\iv\uparrow}-\frac{1}{2}\right) 
\left(n^{f}_{\iv\downarrow}-\frac{1}{2}\right)
%% + \sum_{\iv,\sigma} \epsilon^{f}_{\iv} n^{f}_{\iv\sigma}
 \,\, ,
\end{align}	
where the sums run over a three-dimensional simple cubic lattice, with $\langle{\bf i,j} \rangle$ denoting nearest-neighbor sites, and $\alpha=d$ or $f$. The first term on the r.h.s.~of Eq.\eqref{eq:hamil} corresponds to the hopping of conduction \textit{d}-electrons, while the last one describes the Coulomb repulsion on localized $f$-orbitals. The hybridization between these two orbitals is given by an on-site inter-orbital hopping $V$ and the filling of the lattice is set by adjusting the chemical potential $\mu$. Hereafter, we set the energy scale as being units of the hopping integral, $t$, and define $U_f/t = 6$.  
We have assumed there is no difference in the single particle $d$ and $f$ levels,
that is, $\epsilon_f-\epsilon_d=0$.

We investigate the thermodynamic properties of the 3D PAM by performing finite-temperature auxiliary-field QMC simulations; namely, determinant QMC\cite{blankenbecler81,hirsch85,white89}. The DQMC method is an unbiased technique commonly used to investigate tight-binding Hamiltonians: it maps a $d$-dimensional interacting system to a noninteracting ($d$+1)-dimensional one, with the additional imaginary-time coordinate $0 \leq \tau \leq \beta$, where $\beta$ is the inverse temperature.
Within this approach, one separates the one-body ($\hat {\mathcal K}$) and two-body ($\hat {\mathcal P}$) pieces in the partition function by  using the Trotter-Suzuki decomposition, i.e.~by defining $\beta = L_{\tau} \Delta \tau$, with $L_{\tau}$ being the  number of imaginary-time slices, and $\Delta \tau$ the discretization grid. Then 
\begin{align}
	{\cal Z} &= {\rm Tr}\, e^{-\beta \hat {\mathcal H} } = {\rm Tr}\, \big[ \big(e^{-\Delta\tau ( \hat {\mathcal K} + \hat {\mathcal P})}\big)^{L_{\tau}} \big]\nonumber\\
	 &\approx {\rm Tr}\, \big[ e^{-\Delta\tau \hat
{\mathcal K}} e^{-\Delta\tau \hat {\mathcal P}} e^{-\Delta\tau \hat
{\mathcal K}} e^{-\Delta\tau \hat {\mathcal P}} \cdots \big], 
\end{align}
with an
error proportional to $(\Delta \tau)^2$, but being exact in the limit $\Delta \tau \to 0 $.  
The resulting partition function is rewritten in quadratic (single-body) form through a discrete Hubbard-Stratonovich
 transformation (HST) on the two-body terms, $e^{-\Delta \tau \hat{\mathcal P}}$.  
This HST introduces discrete auxiliary fields with components on each of the space and imaginary-time lattice coordinates,
 which are sampled by Monte Carlo techniques. In this work we choose $\Delta \tau=0.1$, so that the error from the
 Trotter-Suzuki decomposition is less than, or comparable to, statistical errors from the Monte Carlo sampling.
 More details about the method are discussed in Refs.\,\onlinecite{dosSantos03b,gubernatis16,becca17}, and references therein.

Although DQMC is unbiased, its low-temperature application is restricted to systems with particle-hole or other symmetries, owing to the minus-sign problem \cite{Loh90,troyer05,mondaini22}.
 For this reason, our focus is on half-filling,
 $\mu = 0$, where the sign problem is absent.
 Fortunately, this density is of considerable interest, both because of the strong magnetic order favored by commensurate filling, and by the materials for which half filling is appropriate (e.g., the undoped parent compounds of the cuprate superconductors). 

In order to examine signatures for the formation of local momentum and antiferromagnetic long-range order,
 we first investigate the thermodynamic properties, namely, the internal energy and specific heat per site, 
 %% respectively. 
\begin{equation}
    e(T,V)= \frac{1}{N}\langle \hat \calH \rangle~,
    \label{Eq. total_energy}
\end{equation}
\begin{equation}
    c(T)=\frac{1}{N}\frac{d \langle \hat \calH \rangle}{dT}~,
    \,\,
    \label{Eq. specific_heat}
\end{equation}
respectively.
We also investigate the magnetic properties by performing calculations of the antiferromagnetic spin structure factor,
 defined as the Fourier transform of the spin-spin correlation functions for $f$ orbitals,
\begin{equation}
    S_{ff}(\boldsymbol{q}) = \frac{1}{3 N}\sum_{\mathbf{i}, \mathbf{j}} e^{i \mathbf{q}\cdot(\mathbf{r}_i - \mathbf{r}_j)} \, \langle  \Vec S_{f\iv} \cdot \vec S_{f\jv} \rangle ,
    \label{Eq. Structure}
\end{equation}
with  $\boldsymbol{q} = (\pi,\pi,\pi)$.
For singlet formation, we examine a correlator function
\begin{equation}
    C_{\textit{fd}} =  \frac{1}{3 N} \sum_{\iv } \langle \Vec S_{f \iv} \cdot \vec S_{d \iv} \rangle.
    \label{Eq. Singlet}
\end{equation}
Here, we define the fermionic spin operators as
 $\vec S_{f\iv} = (f^{\dagger}_{\uparrow \iv} \,\, f^{\dagger}_{\downarrow \iv}) \, \vec \sigma \, \begin{pmatrix} f_{\uparrow \iv} \\ f_{\downarrow \iv} \end{pmatrix}$,
 with $\vec \sigma$ being the Pauli spin matrices, while the 1/3 factor in Eqs.\,\eqref{Eq. Structure} and \eqref{Eq. Singlet} is for the average value along one spin component.
 
%% This allows us to probe their 
We extract critical behavior by means of the antiferromagnetic correlation ratio
\begin{equation}
    R_{c}(L)= 1-\frac{S_{ff}(\textbf{q}+\delta \textbf{q})}{S_{ff}({\textbf{q}})}
    \label{Eq. correlation_ratio}
\end{equation}
with $|\delta \textbf{q}| = \frac{2\pi}{L}$. 
This is a renormalization-group invariant observable, which crossing points for different lattice sizes provides the critical points\,\cite{Kaul15,Darmawan18}.

In addition to these equal-time correlation functions, we also calculate appropriate unequal-time
 quantities, including the magnetic susceptibility
 \begin{equation}
\chi(\mathbf{q}) = \chi_{dd}(\mathbf{q}) + 2 \chi_{df}(\mathbf{q}) + \chi_{ff}(\mathbf{q}),
 \end{equation}
 with
\begin{equation}
    \chi_{\alpha \gamma} (\bold{q})= \frac{1}{3 N} \sum_{\mathbf{i}, \mathbf{j}} \int_{0}^{\beta}d\tau \bigg\langle {\vec S}_{\alpha, \mathbf{j}}(\tau) \cdot {\vec S}_{\gamma,\mathbf{i}}(0)\bigg\rangle e^{i \mathbf{q}\cdot(\mathbf{r}_i - \mathbf{r}_j)}\,,
    \label{Eq. susceptibillity}
\end{equation}
($\alpha, \gamma=d$, or $f$) where we examine the uniform and staggered cases, $\boldsymbol{q}=(0,0,0)$ and $(\pi,\pi,\pi)$,
 respectively.

\begin{figure}[t]
\includegraphics[scale=0.60]{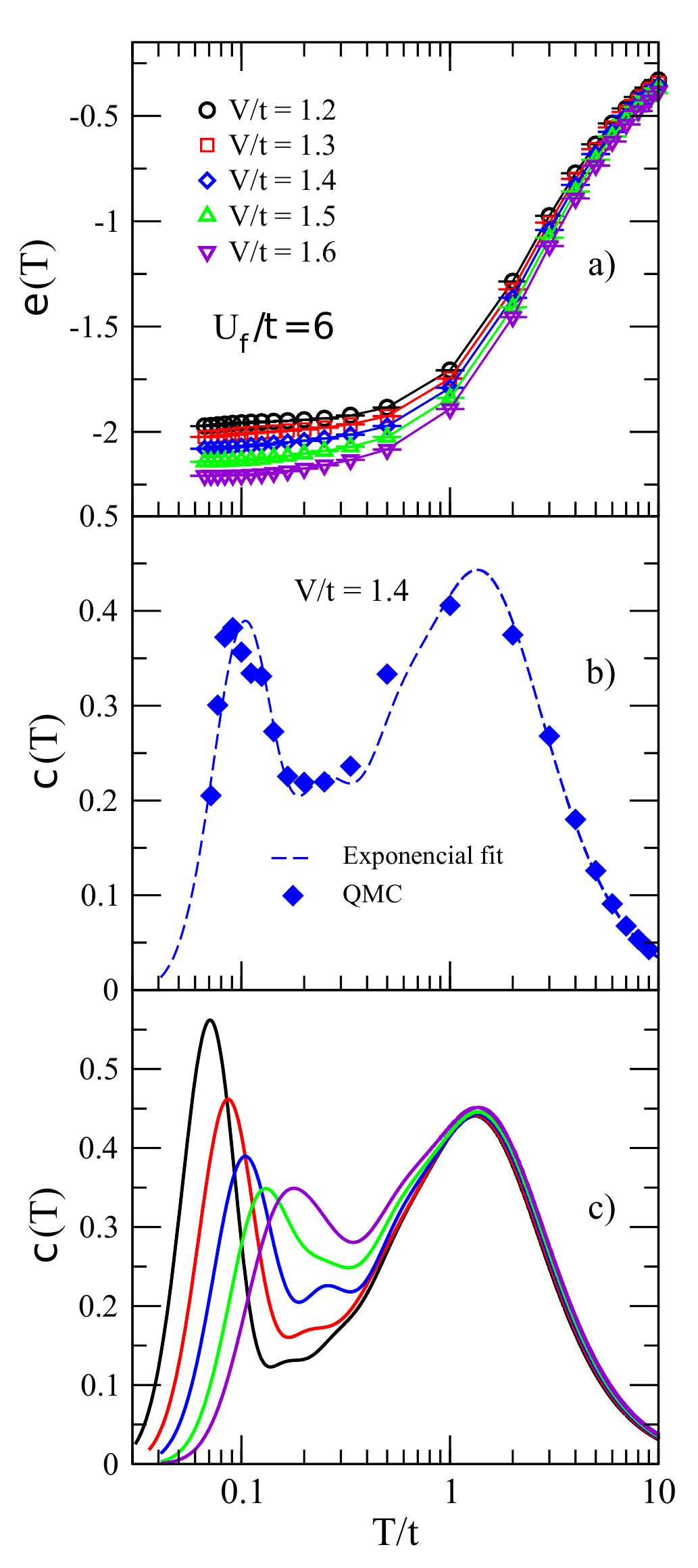} 
\caption{(Color online) (a) Energy as a function of temperature for different $V/t$, and the (b) specific heat as a function of the temperature for $V/t=1.4$; closed symbols represent data from numerical differentiating the energy values, whereas dashed line represents the differentiation of an exponential fit to the energy \cite{paiva01}. (c) Specific heat from  the fit procedure for the same values of $V/t$ as in (a). All data are for $6 \times 6 \times 6$ lattices.}
\label{specific_heat}
\end{figure}

Finally, we also evaluate Nuclear Magnetic Resonance (NMR) quantities, such as the time relaxation rate for the $f$-electrons.
For the low-frequency limit of the dynamic susceptibility, it is defined as (see, e.g.\,Ref.\,\onlinecite{curro09})
\begin{equation}
T_{1,ff}^{-1} = \gamma^2 k_{B} T \lim_{\omega \to 0} \sum_{\mathbf{q}} A^{2}(\mathbf{q})\, \frac{\chi_{ff}^{''}(\mathbf{q},\gamma) }{\hbar \omega},
\label{Eq. NMR}
\end{equation}
where $ A^{2}(\mathbf{q}) $ is the square of the Fourier transform of the hyperfine interaction, and
$\gamma$ is the gyromagnetic ratio.  The latter is related to the nuclear magnetic moment by
 $\gamma \hbar = g \mu_{N} \sqrt{I(I+1)}$, with $\mu_{N}$ being the nuclear magneton, $g$ the nuclear
 $g$-factor, and $I$ the nuclear spin.
$T_{1,ff}^{-1}$ quantifies a characteristic time in which a component of the nuclear spin (of a given $f$-orbital)
%\tcr{{\bf Richard:}  Since we have two types of sites, $d$ and $f$, the reader might wonder which one we are computing, or whether we are getting an average.}
 reaches equilibrium after an external perturbation (magnetic field pulse). It is a dynamical (real frequency)
 quantity whose numerical evaluation usually requires an analytic continuation of the imaginary-time
 DQMC data. Instead, we follow Ref.~\onlinecite{Randeria92}, which performs an approximation to this procedure, leading to
\begin{align}
\label{eq:spin-relaxation2}
\frac{1}{T_{1} T}  = \frac{1}{\pi^2 T^2} \frac{1}{N} \sum_{\mathbf{i}} \bigg\langle S_{f,\mathbf{i}}(\tau=\beta/2) S_{f,\mathbf{i}}(0) \bigg\rangle~.
\end{align}

\section{Results}\label{sec:results}

%%%%%%%%%%%%%%%%%%%%%%%%%%%%%%%%%%%%%%%%%%%%%%%%%%%%%%%%%%%%%%%%%%%%%%%%%%%%%%%%%%%%%%%%%%%%%%%%%%%%%%%%%%%%%%%%%%%%%%
%%%%%%%%%%%%%%%%%%%%%%%%%%%%%%%%%%%%%%%%%%%   Specif Heat   %%%%%%%%%%%%%%%%%%%%%%%%%%%%%%%%%%%%%%%%%%%%%%%%%%%%%%%%%%
%%%%%%%%%%%%%%%%%%%%%%%%%%%%%%%%%%%%%%%%%%%%%%%%%%%%%%%%%%%%%%%%%%%%%%%%%%%%%%%%%%%%%%%%%%%%%%%%%%%%%%%%%%%%%%%%%%%%%

\subsection{Thermodynamic properties}

We start our analysis by discussing the thermodynamic properties of the system -- unless otherwise indicated, the following results are obtained for $6 \times 6 \times 6$ lattice (i.e., 432 sites).
First, we investigate the internal energy, Eq.\,\eqref{Eq. total_energy}, displayed as a function of $T/t$ in Fig.\,\ref{specific_heat}\,(a). 
The derivative of the internal energy with respect to the temperature yields the specific heat, Eq.\,\eqref{Eq. specific_heat}, whose behavior should indicate the occurrence of phase transitions.
Figure \ref{specific_heat}\,(b) displays $c(T)$ for $V/t = 1.4$, exhibiting a two-peak structure.
Since numerical differentiation for a few data points is usually noisy, one may perform a nonlinear fit of the QMC energy points, and differentiate the fitted curve, as shown by the dashed line in Fig.\,\ref{specific_heat}\,(b).
Here we define such a fit on an exponential basis\cite{paiva01}, by the function $e_{\rm fit}(T)=a_{0}+\sum_{n=1}^{M} a_{n}{\rm exp}\,(-\beta n \Delta)$, with a cut-off in $M$ (fixing $M \le 10 $ is enough).
%% Due to 
The agreement indicates the rubustness of the fitting approach, which we repeat 
%% the latter (exponential fit) 
for other values of hybridization  $V/t$, whose results are shown in Fig.\,\ref{specific_heat}\,(c).

\begin{figure}[t]
\includegraphics[scale=0.43]{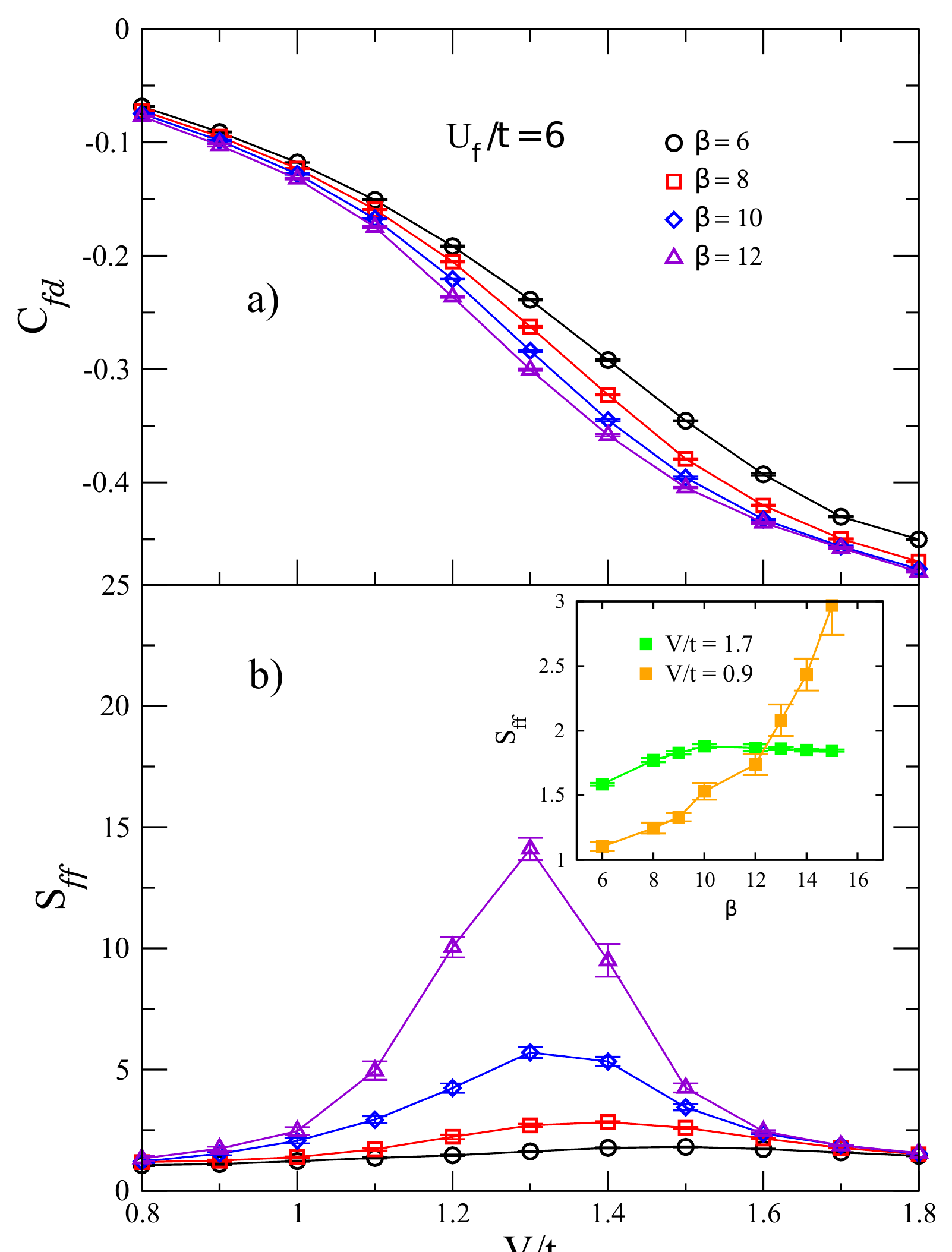} 
\caption{(Color online) (a) Singlet correlator and (b) antiferromagnetic structure factor as functions of the hybridization for different temperatures for $6 \times 6 \times 6$ lattices. The inset in (b) shows the antiferromagnetic structure factor as a function of inverse temperature for hybridizations $V/t=0.9$ and $V/t=1.7$.}
\label{correlator_and_structure}
\end{figure}

%% At this point, 
There are several important features concerning the behavior of $c(T)$ to be emphasized.
First, 
%% one may notice 
a high-temperature hybridization-independent peak at $T/t \sim 1.45$ is evident, which is due to the formation of local moments\,\cite{paiva01}.
We also find a low-temperature peak, that is pushed to higher temperatures  and decreases in intensity as $V/t$ increases.
This low-temperature peak is also seen for the two and three-dimensional Hubbard model\,\cite{Duffy97,paiva01,Muramatsu00}, and usually is associated to collective spin-wave excitations.
In particular, for the three-dimensional Hubbard model this low-temperature peak position signals the N\'eel temperature since the ground state is AFM.
For the 3D PAM, on the other hand, previous calculations for the specific heat were unable to resolve the low-$T$ peak for values of $V/t$ below the AF-singlet quantum critical point~\cite{McMahan98,Huscroft99},
as we have done in this paper.
%% Here, we show well-formed low-$T$ peaks for smaller values of $V/t$ that evolve into a high-$T$ shoulder-like feature as $V/t$ increases.
As one expects a phase transition from the antiferromagnetic to the singlet phase, the position of the low-$T$ peak may not 
%% be a thorough quantity to
uniquely identify the N\'eel temperature for the 3D PAM.
Therefore, to probe whether the low-$T$ peaks indicate the presence of long-range spin correlations, we next investigate the magnetic properties of the system.

%%%%%%%%%%%%%%%%%%%%%%%%%%%%%%%%%%%%%%%%%%%%%%%%%%%%%%%%%%%%%%%%%%%%%%%%%%%%%%%%%%%%%%%%%%%%%%%%%%%%%%%%%%%%%%%%%%%%%%
%%%%%%%%%%%%%%%%%%%%%%%%%%%   Singlet Correlator and Structure Factor%%%%%%%%%%%%%%%%%%%%%%%%%%%%%%%%%%%%%%%%%%%%%%%%%
%%%%%%%%%%%%%%%%%%%%%%%%%%%%%%%%%%%%%%%%%%%%%%%%%%%%%%%%%%%%%%%%%%%%%%%%%%%%%%%%%%%%%%%%%%%%%%%%%%%%%%%%%%%%%%%%%%%%%%
\subsection{Magnetic properties}

We proceed to examine the singlet correlator, $C_{fd}$: a measure of the on-site spin-spin correlation between an \textit{f}-moment and the spin of a \textit{d}-electron. 
We find this quantity is always negative, pointing to antiferromagnetic coupling and singlet formation that increases in strength as $V/t$ increases.
As displayed in Fig.~\ref{correlator_and_structure}(a), $C_{fd}$ has a rapid increase in modulus around $V/t \approx 1.3$, providing an
initial indication of a change 
%% in the regime 
into a singlet phase.

 In order to go beyond this rough indication from the singlet correlator, we also analyze the $f$-orbital AFM structure factor, Eq.~\eqref{Eq. Structure}, shown in
 Fig.~\ref{correlator_and_structure}(b).
 $S_{ff}$ grows rapidly as $V/t$ decreases from $V/t \sim 2$, especially for larger $\beta$, indicating the formation of long range AF order on the $f$ orbitals as the singlet
 correlations diminish (Fig.~\ref{correlator_and_structure}(a)).
 However, after reaching a maximum at $V/t \sim 1.3$, $S_{ff}$ begins falling.  Naively, the $f$-orbital magnetic behavior is symmetric 
 about the maximum.  However, examining the temperature dependence of $S_{ff}$ reveals a profoundly different story.
Specifically, for $V/t=0.9$, $S_{ff}$ while seemingly small, grows very rapidly as $\beta$ increases -- see the inset of Fig.~\ref{correlator_and_structure}(b).
This is in sharp contrast to the behavior at $V/t=1/7$, where $S_{ff}$ saturates at a small value as the temperature is lowered (increasing $\beta$), as also shown in the inset of Fig.~\ref{correlator_and_structure}(b).
This, together with the fact that the singlet correlator is small for $V/t \sim 0.9$ suggests that the low $V/t$ region still has AF order, but that the transition
temperature $T_{N}$ is below the smallest temperature simulated $T_{N} < t/12$.
 
%% Notice that, $S_{ff}$ is suppressed for larger hybridization, and saturates at a small value as the temperature is lowered (increasing $\beta$), as shown in the inset of Fig.~\ref{correlator_and_structure}(b) for $V/t=1.7$.
%% The strong singlet formation (indicated by the $C_{fd}$ behavior) with a small inverse temperature-saturated AFM structure factor are indicators of the nonmagnetic singlet phase at large $V/t$.  

%% On the other hand, for $V/t=0.9$, $S_{ff}$ while seemingly small, grows very rapidly as $\beta$ increases -- see the inset of Fig.~\ref{correlator_and_structure}(b).
%% for $V/t=0.9$.
%% For these fixed non-zero temperatures, $S_{ff}$ exhibits a maximum at $V/t \sim 1.3$ which reflects the non-monotonic nature of the energy scale for antiferromagnetic correlations and, eventually, the N\'eel temperature ($T_N$) expectation.
%% As $V/t$ increases above $V/t \sim 1.3$ the structure factor is drastically reduced.
%% We recall that such behavior for $T_N$ is somehow expected from the well-known Doniach's phase diagram.

\begin{figure}[t]
\includegraphics[scale=0.5]{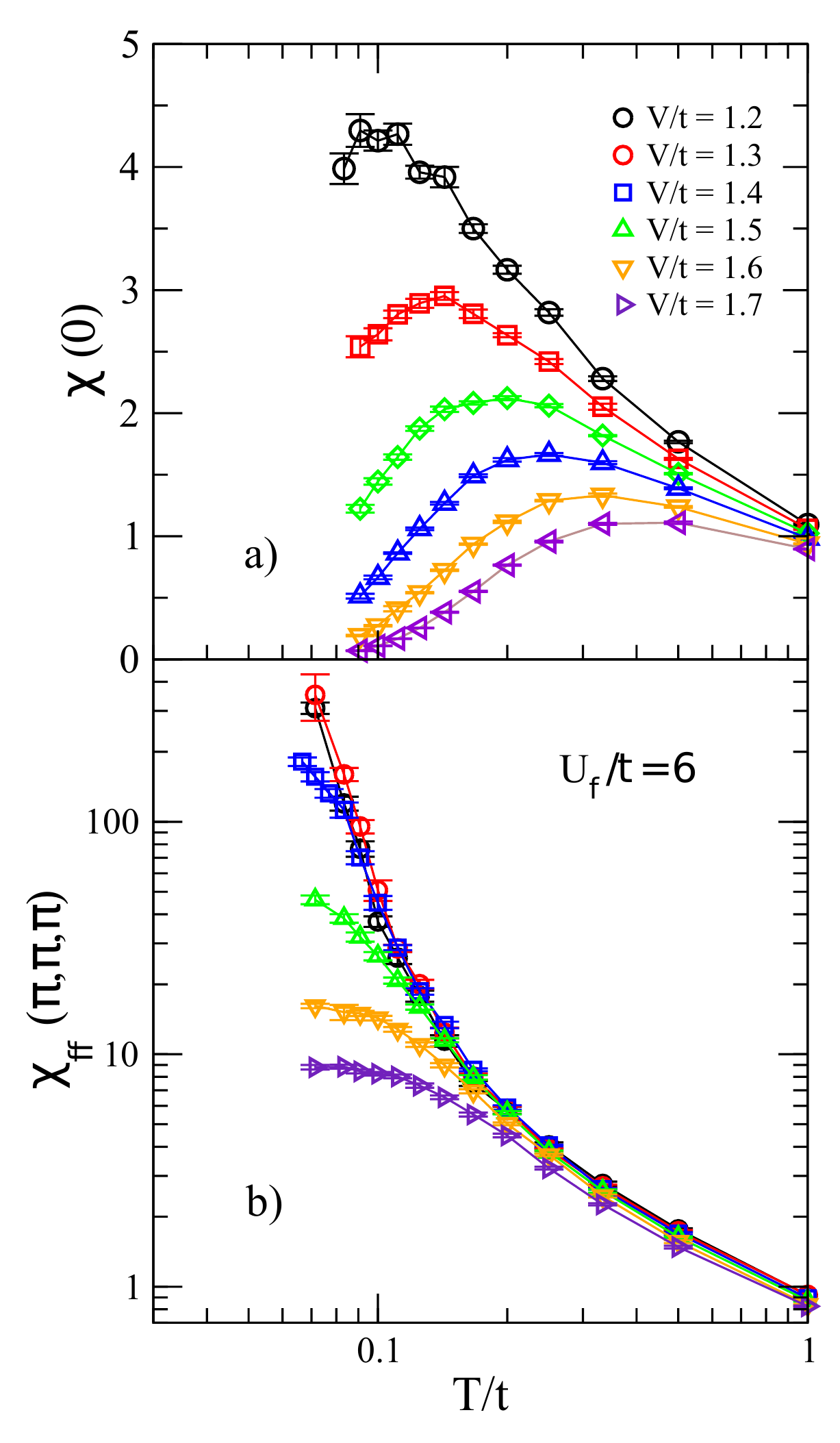}
\caption{(Color online) Total uniform susceptibility (a) and  staggered susceptibility in the $\textit{f}$ sites (b) for different values of $V/t$ on $6 \times 6 \times 6$ lattices.}
\label{fig:suscep}
\end{figure}

%%%%%%%%%%%%%%%%%%%%%%%%%%%%%%%%%%%%%%%%%%%%%%%%%%%%%%%%%%%%%%%%%%%%%%%%%%%%%%%%%%%%%%%%%%%%%%%%%%%%%%%%%%%%%%%%%%%%%%
%%%%%%%%%%%%%%%%%%%%%%%%%%%%%%%%%%%%%   Susceptibility %%%%%%%%%%%%%%%%%%%%%%%%%%%%%%%%%%%%%%%%%%%%%%%%%%%%%%%%%%%%%%%
%%%%%%%%%%%%%%%%%%%%%%%%%%%%%%%%%%%%%%%%%%%%%%%%%%%%%%%%%%%%%%%%%%%%%%%%%%%%%%%%%%%%%%%%%%%%%%%%%%%%%%%%%%%%%%%%%%%%%%

%\subsection{Spin susceptibility and spin gap}

To further understand the magnetic behavior we turn to 
the uniform and staggered magnetic susceptibilities, shown in Figs.~\ref{fig:suscep}(a) and (b), respectively.
For high temperatures, both uniform and staggered susceptibilities show the expected $1/T$ Curie behavior of free spins, while for low temperatures and small hybridizations (e.g., $V/t=1.2$ and 1.3), the uniform susceptibility $\chi(0)$ displays a clear peak, followed by a strong suppression as $T$ decreases.
We recall that, in antiferromagnetic systems, such a peak of $\chi(0)$ occurs at the N\'eel temperature.
Here, as our system may exhibit both AFM and singlet phases, the peak turns into a cusp pushed to higher temperatures for larger $V/t$, blurring the identification of $T_N$.

As will be further discussed later, such a cusp in $\chi(0)$ occurs due to the emergence of the spin-singlet phase, and gives the energy scale for the spin gap formation, $\Delta_{s}$.
Such a gap in the spin excitations is clearly noticed when analyzing the staggered susceptibility, $\chi_{ff}(\pi, \pi, \pi)$, displayed in Fig.~\ref{fig:suscep}(b).
In the thermodynamic limit, the staggered susceptibility should diverge at $T_N$, while $\chi_{ff}(\pi, \pi, \pi) \to 0$ in presence of a spin gap.
Indeed, as shown in Fig.~\ref{fig:suscep}(b) the susceptibility displays a fast increase with decreasing $T/t$ for $V/t  \lesssim 1.4$, while is strongly suppressed for larger hybridizations.

\begin{figure}[t]
\includegraphics[scale=0.6]{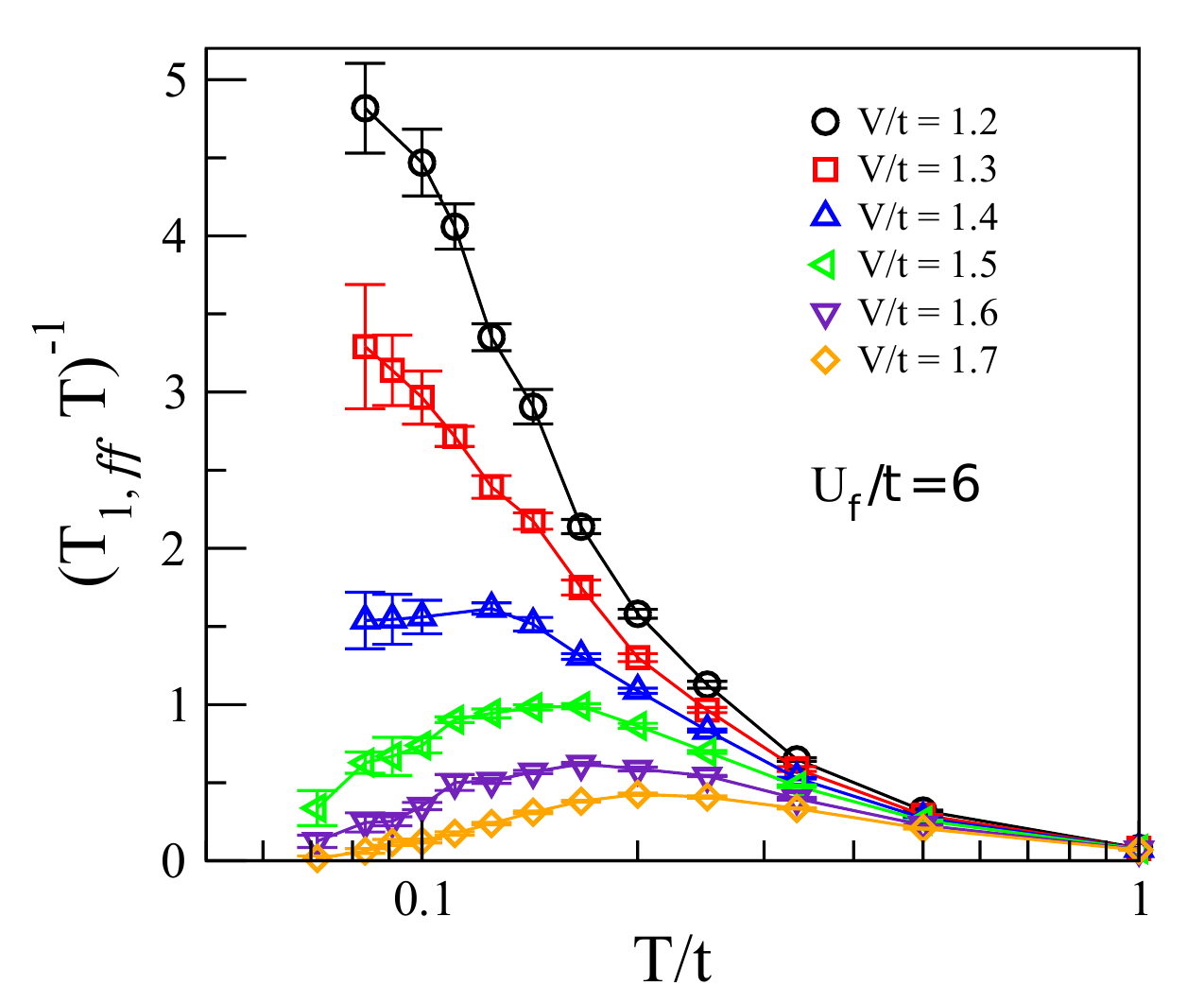} 
\caption{(Color online) Spin-lattice relaxation rate of the 3D PAM as a function of temperature,
 for different hybridizations and $6 \times 6 \times 6$ lattices.}
\label{NMR}
\end{figure}

Early DQMC studies of two-dimensional PAM have shown evidence of a QCP at $V_{c}/t \approx 1.0$, separating an antiferromagnetically (AFM) ordered ground state from a spin-singlet\,\cite{Vekic95}.
The problem was recently revisited by DQMC and Dynamical Vertex Approximation analyses, which show a similar value for the QCP\,\cite{hu17,Schafer19}.
%% Although at ground state, 
One would expect that some finite temperature quantities should reflect the characteristics of the different low $T$ phases,
%% as $T/t$ decreases, 
even for the 2D case (where the occurrence of long-range ordered AFM phase is forbidden at finite temperatures).
Indeed, this is the case of NMR measurements, in particular the spin-lattice relaxation rate, Eq.\,\eqref{Eq. NMR}, which may signal the existence of a spin gap\,\cite{mendes-santos17}.
Previous DQMC studies of the 2D PAM have shown that, at the critical point $V_c$, the $1/T_1 T$ is almost constant at low temperatures, while increasing (reducing) for $V < V_c$ ($V > V_c$)\cite{Costa19}.
Figure \ref{NMR} displays the behavior of the relaxation rate as a function of temperature for different values of $V/t$, for the 3D PAM (linear size $L=6$).
Notice that, within the AFM phase ($V/t =1.2$ or $1.3$) $1/T_1 T$ approaches a finite nonzero value as $T/t \to 0$, consistent with the absence of a spin gap, i.e.\,the presence of spin-wave excitations.
On the other hand, for larger $V$, $1/T_1 T$ decreases monotonically when $T$ is lowered, reflecting a spin-gapped ground state.
Interestingly, the change in behavior of $1/T_1 T$ occurs around $V/t \sim 1.4$, where both results from specific heat and susceptibilities support the disappearance of AFM order.

\begin{figure}[t]
\includegraphics[scale=0.4]{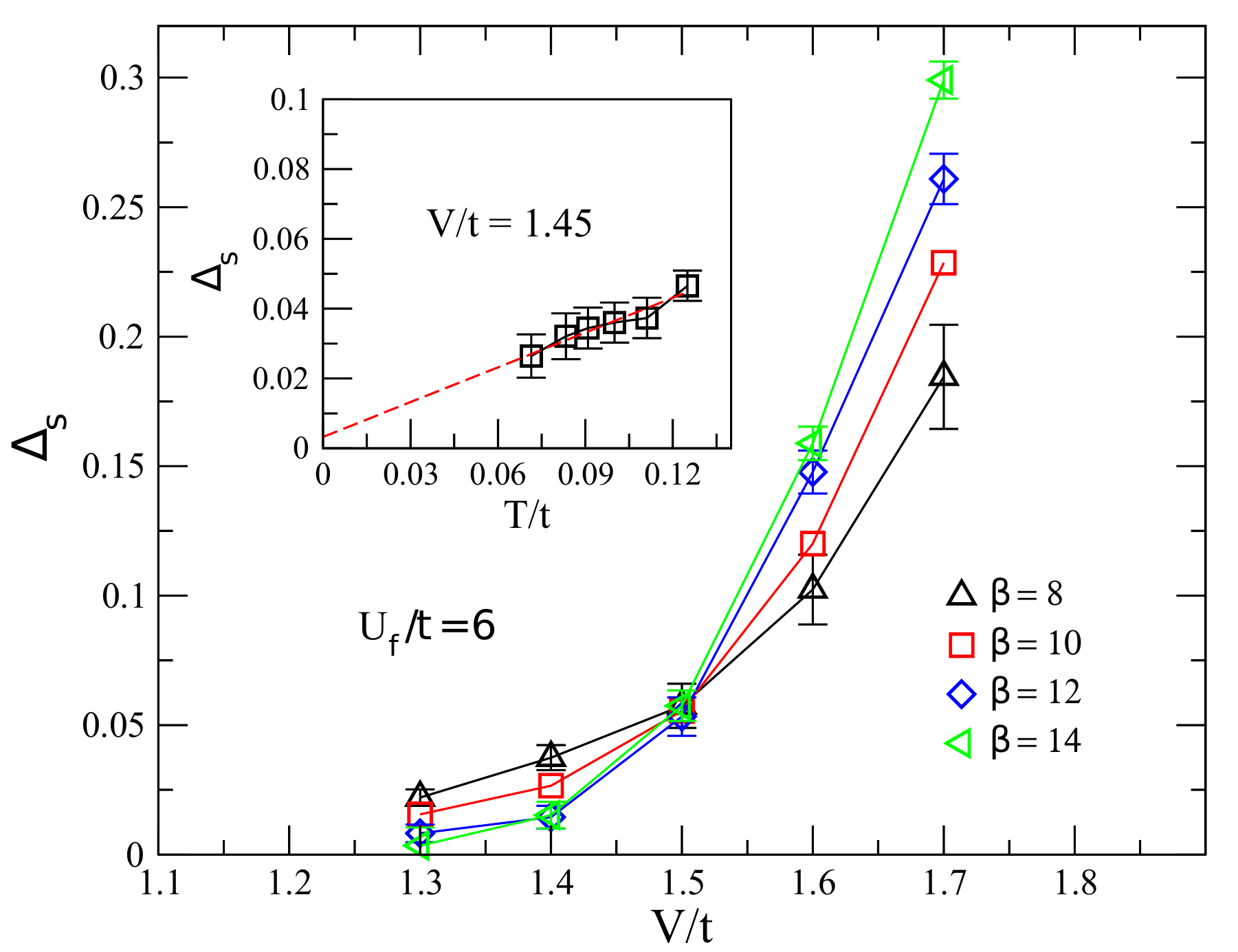} 
\caption{(Color online) 
%The single-particle gap $\Delta_{sp}$ extracted from the long time behavior of the Green’s functions $G(\textbf{k}, \tau$) at $\mathbf{q} = (\pi,\pi,\pi)$ as a function of hybridization. 
The spin gap $\Delta_s$ as a function of hybridization for different inverse temperatures.
As is apparent, there are two distinct behaviors as we move from high to low hybridization, crossing a temperature-independent critical region around $V/t \sim$ 1.5. Inset: The spin-gap
 as a function of temperature for $V/t=1.45$.}
\label{gap}
\end{figure}

In order to estimate and track the spin-gap $\Delta_{s}$ opening, it is useful to analyze the unequal-time spin-spin correlation functions, $C(\mathbf{i},\mathbf{j},\tau) = \langle S^{z}_{\mathbf{j}}(\tau) S^{z}_{\mathbf{i}}(0) \rangle$.
In particular, in the presence of a spin-gap, the asymptotic behavior of the unequal-time staggered structure factor becomes
\begin{align}
\lim_{\tau \to \infty} \frac{1}{N}\sum_{j} (-1)^{|\mathbf{i}- \mathbf{j}|} C(\mathbf{i}=0,\mathbf{j},\tau) \propto \exp(-\Delta_{s} \tau)~,
\end{align}
from which $\Delta_{s}$ may be extracted.
Following this procedure, Fig.~\ref{gap} exhibits our estimation for $\Delta_{s}$ as a function of the hybridization, and at different temperatures.
For large hybridization values, we notice that the gap increases with decreasing temperature, remaining finite at low-$T$.
For small hybridizations, on the other hand, the gap is small and decreases with decreasing temperature; that is, one has a spin gapless phase, in line with our previous expectation of AFM order.
The change in such behavior occurs for $V/t \approx 1.5$, where the spin-gap is temperature independent; in fact, as shown in the inset of Fig.~\ref{gap}, the extrapolation to $T/t \to 0$ for  $V/t = 1.45$ leads to a vanishing gap. 
In summary, the results presented in Figs.\,\ref{fig:suscep}-\ref{gap} point to a QCP at $V_c/t=1.45 \pm 0.05$, separating an AFM phase at low hybridization from a singlet phase at large $V/t$.

\begin{figure}[t]
\includegraphics[scale=0.4]{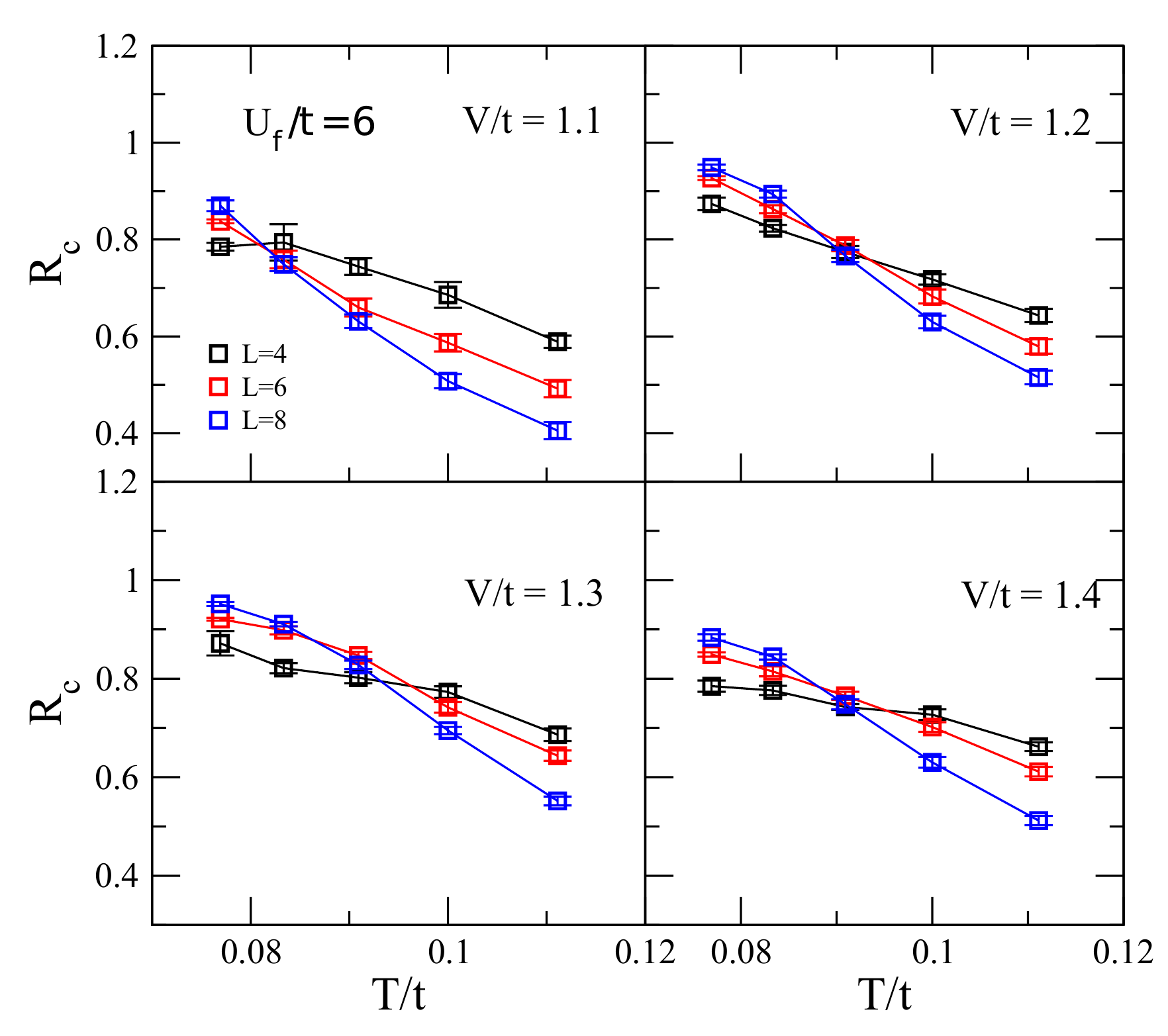} 
\caption{(Color online) Determinant quantum Monte Carlo results for the AF correlation ratio, as function of inverse temperature,
 for different lattice sizes and hybridizations, at fixed  $U_{f}/t=6$. The crossing points
 separate the paramagnetic phase and N\`eel phases.}
\label{Rc}
\end{figure}

Although we have estimated the location of the QCP, the N\'eel temperatures (for $V < V_c$) are ill-defined by the specific heat and homogeneous susceptibility behavior.
A thorough probe of $T_N$ is provided by a finite-size scaling analysis of the correlation ratio, Eq.~\eqref{Eq. correlation_ratio}, a renormalization-group invariant, and shown in Fig.~\ref{Rc} as a function of temperature, for different hybridizations and system sizes.
The crossing of the $Rc(L)$ for different lattice sizes provides the critical temperature.
Since the correlation ratio shows no abrupt features, the paramagnetic metallic to antiferromagnetic insulator finite temperature phase transition is more likely to be continuous.
For the range of $V/t$ analyzed we have $T_N/t \simeq 0.08 - 0.10$.  However, for smaller hybridizations $T_N$ will decrease, as previously discussed in the
context of  Fig.~\ref{correlator_and_structure}(b).
Going to such low $T/t$ is challenging to achieve in three-dimensional geometries.
It is worth noting that this computational limitation also affects the determination of $T_N$ in  the 3D single band Hubbard model,
where $T_N \sim t\, e^{-a \sqrt{t/U}}$ is very small at weak coupling\cite{slater51,hirsch87,scalettar89,nagaosa99,fuchs11,kozik13}.

We conclude by comparing the energy scales of our results with those from the literature.
Early QMC attempts to investigate the three-dimensional PAM were performed in Refs.\,\onlinecite{McMahan98,Huscroft99,Paiva03}, providing hints that such an AFM long-range order requires very low temperatures to be achieved.
By `low-temperature' we mean by comparison with the 3D Hubbard model\,\cite{Muramatsu00,Hirschmeier15,Khatami16}, whose maximum of the N\'eel temperature is $T_{N}/t \approx 0.33$, and occurs at $U_{f}/t \approx 8$.
Here, we provide $T_N$ for the 3D PAM, showing that these energy scales are indeed much lower than those from the Hubbard model\footnote{Interestingly, the N\'eel temperatures obtained by DMFT\,\cite{Held00,Schafer19} are compatible with our QMC ones.}. 
This difference is expected due to the weak indirect coupling between $f$-electrons mediated by conduction ones -- first addressed by DMFT studies\,\cite{Held00}.
In view of this, it is expected that the value of $V_c / t$ should change for different values of $U_f$, but the maxima of $T_N$ should keep similar values as those presented for $U_f / t =6$.

%%%%%%%%%%%%%%%%%%%%%%%%%%%%%%%%%%%%%%%%%%%%%%%%%%%%%%%%%%%%%%%%%%%%%%%%%%%%%%%%%%%%%%%%%%%%%%%%%%%%%%%%%%%%%%%%%%%%%%
%%%%%%%%%%%%%%%%%%%%%%%%%%%%%%%%%%%%% DOS %%%%%%%%%%%%%%%%%%%%%%%%%%%%%%%%%%%%%%%%%%%%%%%%%%%%%%%%%%%%%%%%%%%%%%%
%%%%%%%%%%%%%%%%%%%%%%%%%%%%%%%%%%%%%%%%%%%%%%%%%%%%%%%%%%%%%%%%%%%%%%%%%%%%%%%%%%%%%%%%%%%%%%%%%%%%%%%%%%%%%%%%%%%%%
\subsection{Transport properties}

In the ground state, the AFM and singlet phases are both insulators.
However, at finite temperatures, a crossover from a metallic to insulating behavior should occur at an appropriate energy scale.
Therefore, as a complementary study, we also probe such a crossover as the temperature is lowered, showing that its features can be identified from the behavior of thermodynamic quantities\cite{Kim20}.

\begin{figure}[t]
\includegraphics[scale=0.45]{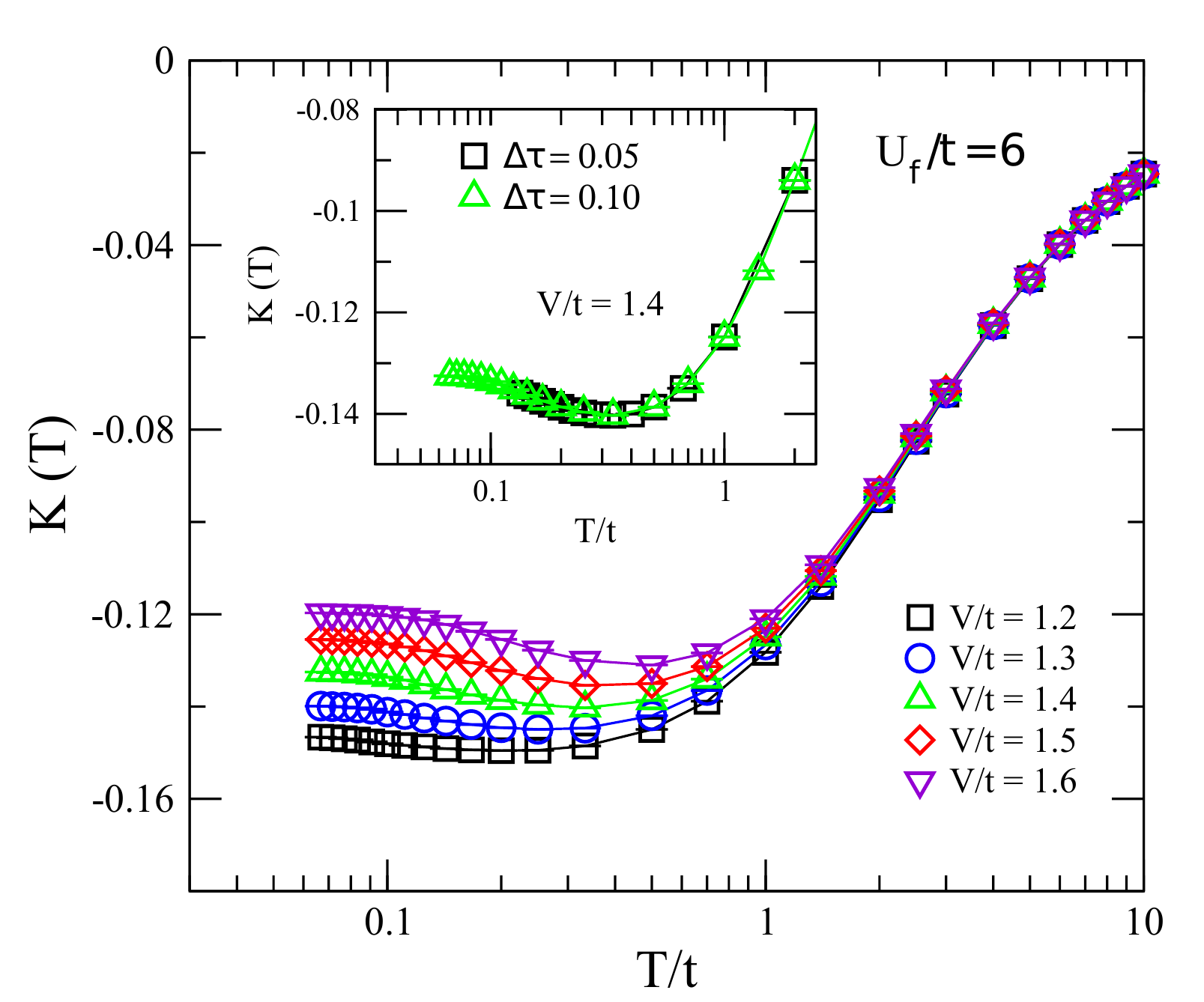} 
\caption{(Color online) Kinetic energy $K$  as a function of temperature for different values of $V/t$ for $6 \times 6 \times 6$ lattices. The inset shows $K$ as a function of $T/t$ for $V/t=1.4$ and two different values of imaginary time discretization: $\Delta \tau =0.05$ (squares) and $\Delta \tau =0.10$ (triangles). 
}
\label{hopping}
\end{figure}

We first examine the $c$-electrons kinetic energy per particle, 
\begin{equation}
\big\langle K \big\rangle =-\frac{t}{N}\bigg\langle \sum_{\langle \mathbf{i}, \mathbf{j} \rangle,\sigma}(d_{\mathbf{i},\sigma}^\dagger d_{\mathbf{j},\sigma} + H.c)\bigg\rangle,
\end{equation}
whose behavior is presented in Fig.~\ref{hopping} as a function of temperature, for different values of hybridization.
For all hybridization values, the following behavior is noticed: at high temperatures, $\big\langle K \big\rangle$ is reduced as temperature is lowered, reaching a minimum for a given energy scale $T_{\rm min(\langle K \rangle)}$ and, therefore, increasing as $T \to 0$ when $ T < T_{\rm min(\langle K \rangle)}$.
The sign change in $\frac{\partial \big\langle K \big\rangle}{\partial T}$, from positive to negative, is consistent with an evolution from metallic to insulator behavior, respectively.
In order to emphasize the absence of imaginary time discretization errors, the inset in Fig.~\ref{hopping} shows the kinetic energy per particle for $V/t=1.4$ as a function of temperature for two different values of $\Delta \tau$, from which is clear that our choice of $\Delta \tau$ does not affect the results of $\big\langle K \big\rangle$.
This result of the kinetic energy (suggesting an insulator for $ T < T_{\rm min(\langle K \rangle)}$), combined with the fact that $\frac{\partial}{\partial T} \chi(0) \leq 0$ at lower temperatures (consistent with a Pauli metal for $d$-electrons + local moment Curie behavior for $f$-electrons, instead of a spin-gapped insulator; see Fig.~\ref{fig:suscep}), point out to a bad metal phase.

\begin{figure}[t]
\includegraphics[scale=0.4]{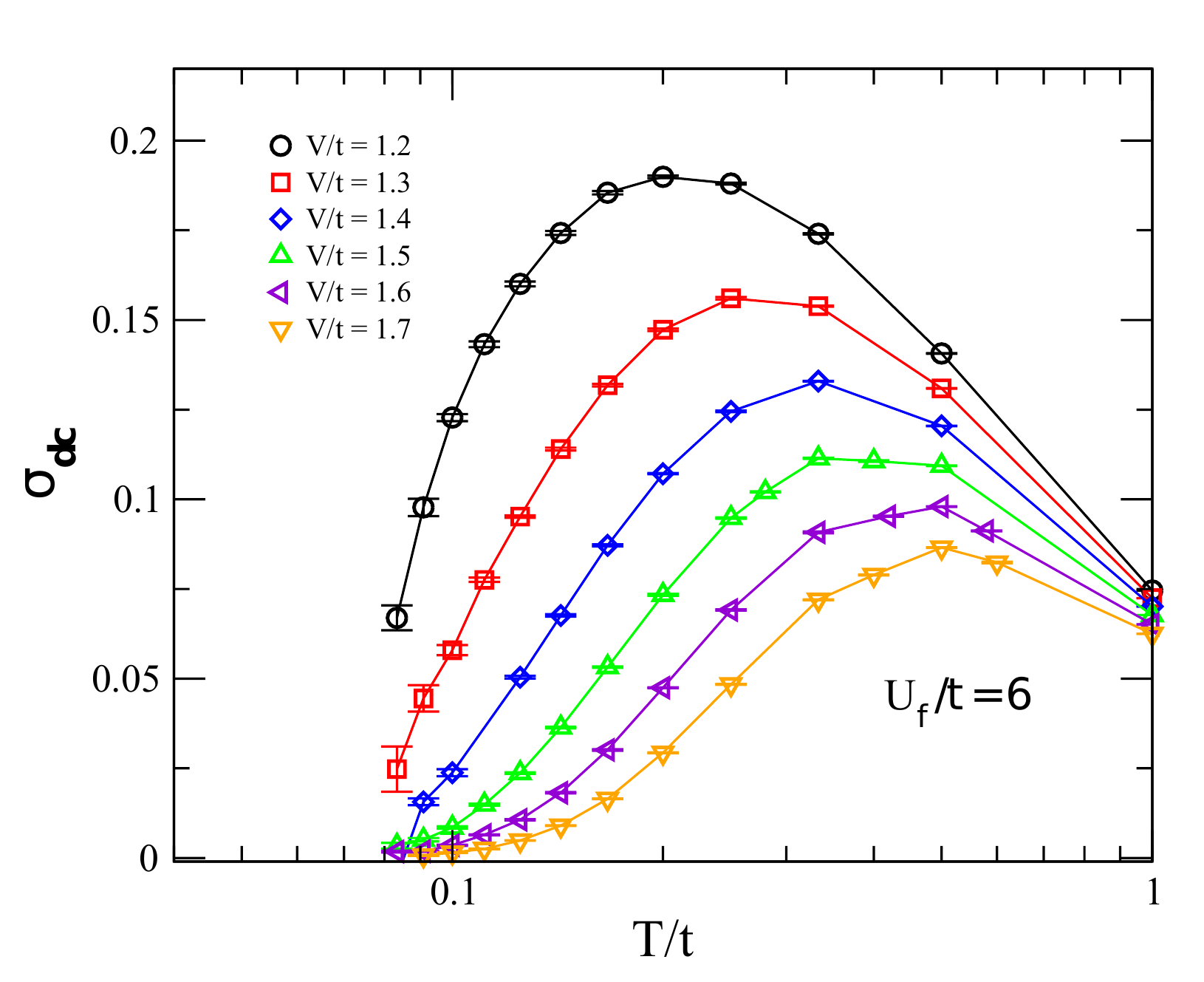} 
\caption{(Color online) $dc$ conductivity as a function of the temperature, for different values of $V/t$ and $6 \times 6 \times 6$ lattices.}
\label{conductivity}
\end{figure}

To further emphasize this bad metallic phase, we also investigate the dc conductivity,
\begin{equation}\label{eq:sigma_dc}
\sigma_{dc} = \frac{\beta^2}{\pi} \Lambda_{xx}(\mathbf{q=0}, \tau = \beta/2),
\end{equation}
in which
\begin{equation}
\Lambda_{xx}(\mathbf{q}, \tau ) = 
\langle j_{x}(\mathbf{q}, \tau) j_{x}(-\mathbf{q}, 0)  \rangle,
\end{equation}
with $ j_{x}(\mathbf{q}, \tau) $ being the Fourier transform of the unequal-time current-current correlation functions
\begin{equation}
j_x(\mathbf{i},\tau)=\mathrm{e}^{\tau\mathcal{H}}
  \left[
        it\sum_\sigma
            \left(d_{\mathbf{i}+\mathbf{x}\sigma}^\dagger 
                  d_{\mathbf{i}\sigma}^{\phantom{\dagger}}
                  - 
                  d_{\mathbf{i}\sigma}^\dagger  
                  d_{\mathbf{i}+\mathbf{x}\sigma}^{\phantom{\dagger}}
            \right)
  \right]
\mathrm{e}^{-\tau\mathcal{H}};
\label{jx}
\end{equation}
see, e.g., Refs.\onlinecite{Trivedi96,Denteneer99,Mondaini12}.
The behavior of $\sigma_{dc}$ as a function of temperature is displayed in Fig.~\ref{conductivity}, which shows $\sigma_{dc} \to 0$ as $T \to 0$ for all $V/t$, consistent with an insulating phase ground state for the entire range of hybridizations.
However, this insulating feature -- i.e.~$\frac{\partial \sigma_{dc}}{\partial T} < 0$ -- changes depending on the temperature, with $\sigma_{dc}$ exhibiting a maximum at $T_{\rm max(\sigma_{dc})}$.
Interestingly, the values of $T_{\rm max(\sigma_{dc})}$ are consistent with $T_{\rm min(\langle K \rangle)}$, both signaling a crossover into a bad metallic phase.

%%%%%%%%%%%%%%%%%%%%%%%%%%%%%%%%%%%%%%%%%%%%%%%%%%%%%%%%%%%%%%%%%%
%%%%%%%%%%%%%%%%%%%%%%%%%%%%%%%%%%%%%%%%%%%%%%%%%%%%%%%%%%%%%%%%%%
\section{Conclusions}
\label{sec:conc}

In this work, we have investigated the thermodynamic, magnetic, and transport properties of the three-dimensional Periodic
Anderson Model through unbiased DQMC simulations.
For the thermodynamic properties, we examined the specific heat for different hybridizations.
We find a broad, high-temperature peak that is independent of $V/t$, while the low-$T$ peak, on the other hand, is strongly $V$-dependent.
This low-$T$ peak is associated with the emergence of the AFM phase for $V < V_c $.  However, it does not give precise information about the singlet phase for $V > V_c $\,\footnote{We expect that, for larger lattice sizes, the low-$T$ peak would be enhanced for $V < V_c$, and suppressed for $V > V_c$, emphasizing the transition.}.

 \begin{figure}[t]
\includegraphics[scale=0.33]{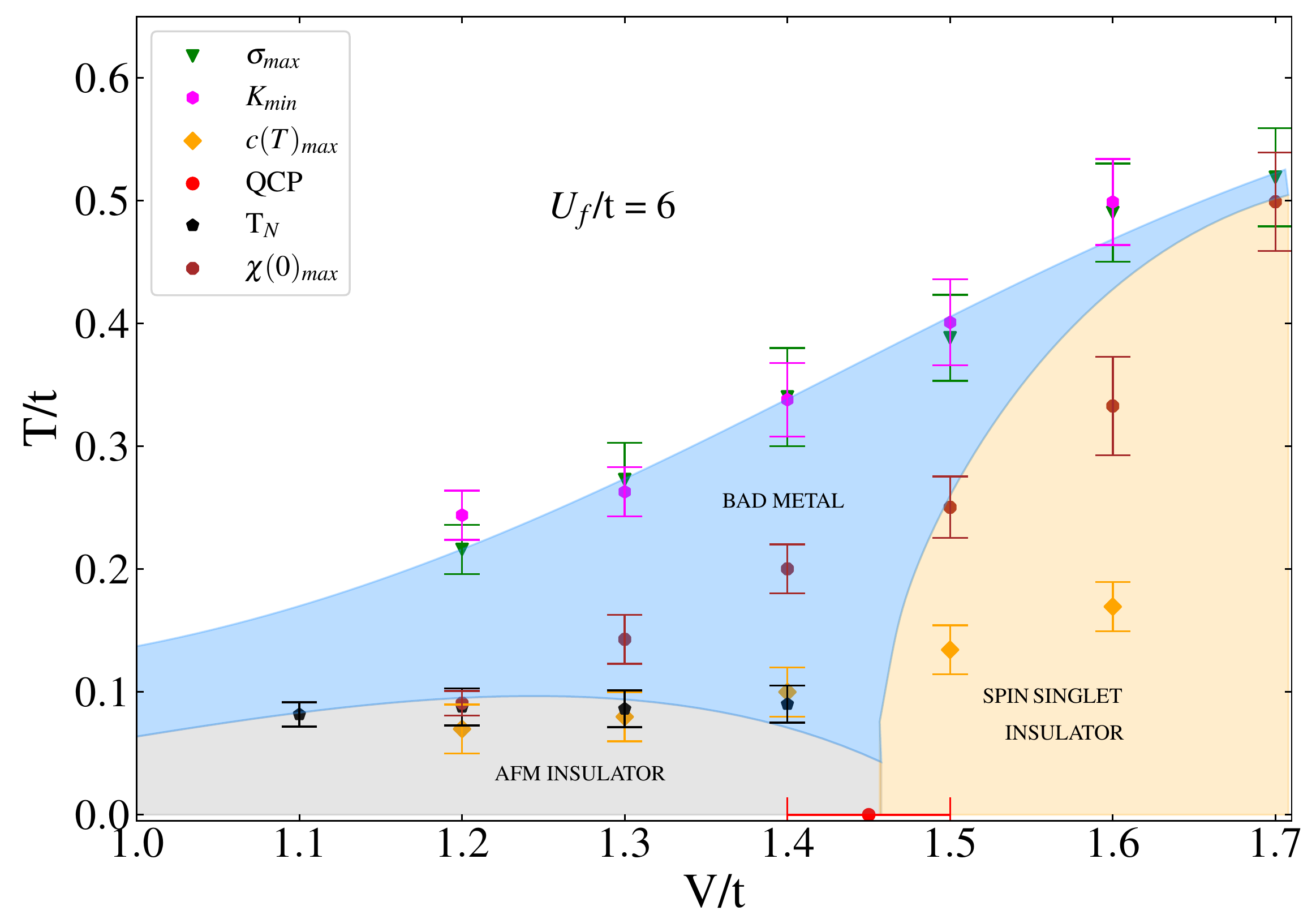} 
\caption{(Color online) Energy scales for the three dimensional Periodic 
Anderson Model for different values of $V/t$. 
 For comparison, the N\'eel temperature $T_N$ of the 3D Hubbard model, when plotted versus
 on-site interaction $U_{f}/t$, exhibits a broad maximum with
$T_N/t \sim 0.3$ in the range $6 \lesssim U_{f}/t \lesssim 10$.}
\label{diagram}
\end{figure}

To get more quantitative insight into the singlet formation,
we investigated the AF structure factor, and
correlator function
%% , which provided hints on the singlet formation and spin-spin correlations 
for different $V/t$ and temperatures.
In particular, the AF structure factor pointed out the energy scale for the quantum critical point.
This quantum phase transition between an AFM state and a singlet state was obtained by analyzing the spin gap, the NMR time relaxation rate, and the staggered magnetic susceptibility, which agree with $V_c/t = 1.45 \pm 0.05$.
The N\'eel temperature within the AFM phase was probed by the antiferromagnetic correlation ratio and lies in the range $T/t \simeq 0.08 - 0.10$ for all $V/t$ investigated. 
Finally, by investigating the transport properties, we find out that there is a region in which the system behaves as a bad metal, i.e.~it presents insulator features for the kinetic energy (namely $\frac{\partial \langle K \rangle}{\partial T} < 0$), while keeping metallic ones for the homogeneous susceptibility (namely $\frac{\partial \chi}{\partial T} \leq 0$ -- a Pauli metal for $d$-electrons + local moment Curie behavior for $f$ ones).

In summary, our work presents detailed finite temperature results for the three-dimensional Periodic Anderson Model in a simple cubic lattice, allowing the quantitative determination of
different energy scales. 
To summarize our findings, Fig.~\ref{diagram} presents these energy scales, and the resulting finite temperature phase diagram.
In particular, we show [i] the quantum critical point, QCP, [ii] the temperature of the maxima for the homogeneous susceptibility, $\chi (0)_{\rm max}$, [iii] the minima for the $d$-electrons hopping energy, $K_{\rm min}$, [iv] the temperature of the maxima of the conductivity $\sigma_{\rm max}$, [v] the low-temperature specific heat peak position $c(T)_{\rm max}$, and finally [vi] the Néel temperatures $T_N$ obtained from a Finite Size Scaling analysis.
Knowing precise values for these energy scales at half-filling is an essential step towards examining other more complex facets of this model which emerge with doping, including the occurrence of $d$-wave superconductivity due to frustration effects\,\cite{Wu15}, or the enhancement of magnetism due to depletion\,\cite{Assaad02,costa18a,Jiang20}.
Indeed, Fig.~\ref{diagram} provides detailed data for the PAM across an interesting part of its phase diagram encompassing the AF-singlet QCP, which may help guide future theoretical and experimental work on heavy fermion materials.

%%%%%%%%%%%%%%%%%%%%%%%%%%%%%%%%%%%%%%%%%%%%%%%%%%%%%%%%%%%%%%%%%%
%%%%%%%%%%%%%%%%%%%%%%%%%%%%%%%%%%%%%%%%%%%%%%%%%%%%%%%%%%%%%%%%%%
\section*{ACKNOWLEDGMENTS}
%%%%%%%%%%%%%%%%%%%%%%%%%%%%%%%%%%%%%%%%%%%%%%%%%%%%%%%%%%%%%%%%%%
%%%%%%%%%%%%%%%%%%%%%%%%%%%%%%%%%%%%%%%%%%%%%%%%%%%%%%%%%%%%%%%%%%

The authors are grateful to the Brazilian Agencies Conselho Nacional de Desenvolvimento Cient\'\i fico e
 Tecnol\'ogico (CNPq), Coordena\c c\~ao de Aperfei\c coamento de Pessoal de Ensino Superior (CAPES),
 Funda\c c\~ao Carlos Chagas de Apoio \`a Pesquisa do Estado do Rio de Janeiro (FAPERJ), and Instituto 
Nacional de Ciência e Tecnologia de Informação Quântica (INCT-IQ) for funding this project.
N.C.C. acknowledges financial support from CNPq, Grant No. 313065/2021-7, and from FAPERJ -- Funda\c{c}\~ao Carlos Chagas Filho de Amparo \`a Pesquisa do Estado do Rio de Janeiro --, Grant No.\,200.258/2023 (SEI-260003/000623/2023). T. P. acknowledges financial support from CNPq Grant Nos. 403130/2021-2 and 308335/2019-8, and FAPERJ Grant Nos. E-26/200.959/2022 and E-26/210.100/2023.
The work of R.S.~was supported by the grant DE-SC-0014671, funded by the U.S. Department of Energy, Office of Science.

%%%%%%%%%%%%%%%%%%%%%%%%%%%%%%%%%%%%%%%%%%%%%%%%%%%%%%%%%%%%%%%%%%%%%%%%
%%%
%%%%%     BIBLIOGRAPHY
%%%%%%%%%%%%%%%%%%%%%%%%%%%%%%%%%%%%%%%%%%%%%%%%%%%%%%%%%%%%%%%%%%%%%%%%%%%

\color{black}

%\begin{thebibliography}{100}

\bibliography{bib_pam.bib}

\begin{thebibliography}{55}%
\makeatletter
\providecommand \@ifxundefined [1]{%
 \@ifx{#1\undefined}
}%
\providecommand \@ifnum [1]{%
 \ifnum #1\expandafter \@firstoftwo
 \else \expandafter \@secondoftwo
 \fi
}%
\providecommand \@ifx [1]{%
 \ifx #1\expandafter \@firstoftwo
 \else \expandafter \@secondoftwo
 \fi
}%
\providecommand \natexlab [1]{#1}%
\providecommand \enquote  [1]{``#1''}%
\providecommand \bibnamefont  [1]{#1}%
\providecommand \bibfnamefont [1]{#1}%
\providecommand \citenamefont [1]{#1}%
\providecommand \href@noop [0]{\@secondoftwo}%
\providecommand \href [0]{\begingroup \@sanitize@url \@href}%
\providecommand \@href[1]{\@@startlink{#1}\@@href}%
\providecommand \@@href[1]{\endgroup#1\@@endlink}%
\providecommand \@sanitize@url [0]{\catcode `\\12\catcode `\$12\catcode
  `\&12\catcode `\#12\catcode `\^12\catcode `\_12\catcode `\%12\relax}%
\providecommand \@@startlink[1]{}%
\providecommand \@@endlink[0]{}%
\providecommand \url  [0]{\begingroup\@sanitize@url \@url }%
\providecommand \@url [1]{\endgroup\@href {#1}{\urlprefix }}%
\providecommand \urlprefix  [0]{URL }%
\providecommand \Eprint [0]{\href }%
\providecommand \doibase [0]{http://dx.doi.org/}%
\providecommand \selectlanguage [0]{\@gobble}%
\providecommand \bibinfo  [0]{\@secondoftwo}%
\providecommand \bibfield  [0]{\@secondoftwo}%
\providecommand \translation [1]{[#1]}%
\providecommand \BibitemOpen [0]{}%
\providecommand \bibitemStop [0]{}%
\providecommand \bibitemNoStop [0]{.\EOS\space}%
\providecommand \EOS [0]{\spacefactor3000\relax}%
\providecommand \BibitemShut  [1]{\csname bibitem#1\endcsname}%
\let\auto@bib@innerbib\@empty
%</preamble>
\bibitem [{\citenamefont {Raczkowski}\ \emph {et~al.}(2020)\citenamefont
  {Raczkowski}, \citenamefont {Peters}, \citenamefont {Ph\`ung}, \citenamefont
  {Takemori}, \citenamefont {Assaad}, \citenamefont {Honecker},\ and\
  \citenamefont {Vahedi}}]{Raczkowski20}%
  \BibitemOpen
  \bibfield  {author} {\bibinfo {author} {\bibfnamefont {Marcin}\ \bibnamefont
  {Raczkowski}}, \bibinfo {author} {\bibfnamefont {Robert}\ \bibnamefont
  {Peters}}, \bibinfo {author} {\bibfnamefont {Th\d{i}~Thu}\ \bibnamefont
  {Ph\`ung}}, \bibinfo {author} {\bibfnamefont {Nayuta}\ \bibnamefont
  {Takemori}}, \bibinfo {author} {\bibfnamefont {Fakher~F.}\ \bibnamefont
  {Assaad}}, \bibinfo {author} {\bibfnamefont {Andreas}\ \bibnamefont
  {Honecker}}, \ and\ \bibinfo {author} {\bibfnamefont {Javad}\ \bibnamefont
  {Vahedi}},\ }\bibfield  {title} {\enquote {\bibinfo {title} {{Hubbard model
  on the honeycomb lattice: From static and dynamical mean-field theories to
  lattice quantum Monte Carlo simulations}},}\ }\href {\doibase
  10.1103/PhysRevB.101.125103} {\bibfield  {journal} {\bibinfo  {journal}
  {Phys. Rev. B}\ }\textbf {\bibinfo {volume} {101}},\ \bibinfo {pages}
  {125103} (\bibinfo {year} {2020})}\BibitemShut {NoStop}%
\bibitem [{\citenamefont {Paleari}\ \emph {et~al.}(2021)\citenamefont
  {Paleari}, \citenamefont {H\'ebert}, \citenamefont {Cohen-Stead},
  \citenamefont {Barros}, \citenamefont {Scalettar},\ and\ \citenamefont
  {Batrouni}}]{Paleari21}%
  \BibitemOpen
  \bibfield  {author} {\bibinfo {author} {\bibfnamefont {G.}~\bibnamefont
  {Paleari}}, \bibinfo {author} {\bibfnamefont {F.}~\bibnamefont {H\'ebert}},
  \bibinfo {author} {\bibfnamefont {B.}~\bibnamefont {Cohen-Stead}}, \bibinfo
  {author} {\bibfnamefont {K.}~\bibnamefont {Barros}}, \bibinfo {author}
  {\bibfnamefont {RT.}\ \bibnamefont {Scalettar}}, \ and\ \bibinfo {author}
  {\bibfnamefont {G.~G.}\ \bibnamefont {Batrouni}},\ }\bibfield  {title}
  {\enquote {\bibinfo {title} {{Quantum Monte Carlo study of an anharmonic
  Holstein model}},}\ }\href {\doibase 10.1103/PhysRevB.103.195117} {\bibfield
  {journal} {\bibinfo  {journal} {Phys. Rev. B}\ }\textbf {\bibinfo {volume}
  {103}},\ \bibinfo {pages} {195117} (\bibinfo {year} {2021})}\BibitemShut
  {NoStop}%
\bibitem [{\citenamefont {Xing}\ \emph {et~al.}(2021)\citenamefont {Xing},
  \citenamefont {Chiu}, \citenamefont {Poletti}, \citenamefont {Scalettar},\
  and\ \citenamefont {Batrouni}}]{Xing21}%
  \BibitemOpen
  \bibfield  {author} {\bibinfo {author} {\bibfnamefont {Bo}~\bibnamefont
  {Xing}}, \bibinfo {author} {\bibfnamefont {Wei-Ting}\ \bibnamefont {Chiu}},
  \bibinfo {author} {\bibfnamefont {Dario}\ \bibnamefont {Poletti}}, \bibinfo
  {author} {\bibfnamefont {R.~T.}\ \bibnamefont {Scalettar}}, \ and\ \bibinfo
  {author} {\bibfnamefont {George}\ \bibnamefont {Batrouni}},\ }\bibfield
  {title} {\enquote {\bibinfo {title} {{Quantum Monte Carlo simulations of the
  2D Su-Schrieffer-Heeger model}},}\ }\href {\doibase
  10.1103/PhysRevLett.126.017601} {\bibfield  {journal} {\bibinfo  {journal}
  {Phys. Rev. Lett.}\ }\textbf {\bibinfo {volume} {126}},\ \bibinfo {pages}
  {017601} (\bibinfo {year} {2021})}\BibitemShut {NoStop}%
\bibitem [{\citenamefont {Hu}\ \emph {et~al.}(2019)\citenamefont {Hu},
  \citenamefont {Dong},\ and\ \citenamefont {Yang}}]{Hu19}%
  \BibitemOpen
  \bibfield  {author} {\bibinfo {author} {\bibfnamefont {Danqing}\ \bibnamefont
  {Hu}}, \bibinfo {author} {\bibfnamefont {Jian-Jun}\ \bibnamefont {Dong}}, \
  and\ \bibinfo {author} {\bibfnamefont {Yi-feng}\ \bibnamefont {Yang}},\
  }\bibfield  {title} {\enquote {\bibinfo {title} {{Hybridization fluctuations
  in the half-filled periodic Anderson model}},}\ }\href {\doibase
  10.1103/PhysRevB.100.195133} {\bibfield  {journal} {\bibinfo  {journal}
  {Phys. Rev. B}\ }\textbf {\bibinfo {volume} {100}},\ \bibinfo {pages}
  {195133} (\bibinfo {year} {2019})}\BibitemShut {NoStop}%
\bibitem [{\citenamefont {Scalapino}(2012)}]{Scalapino12}%
  \BibitemOpen
  \bibfield  {author} {\bibinfo {author} {\bibfnamefont {D.~J.}\ \bibnamefont
  {Scalapino}},\ }\bibfield  {title} {\enquote {\bibinfo {title} {A common
  thread: The pairing interaction for unconventional superconductors},}\ }\href
  {\doibase 10.1103/RevModPhys.84.1383} {\bibfield  {journal} {\bibinfo
  {journal} {Rev. Mod. Phys.}\ }\textbf {\bibinfo {volume} {84}},\ \bibinfo
  {pages} {1383--1417} (\bibinfo {year} {2012})}\BibitemShut {NoStop}%
\bibitem [{\citenamefont {Kotov}\ \emph {et~al.}(2012)\citenamefont {Kotov},
  \citenamefont {Uchoa}, \citenamefont {Pereira}, \citenamefont {Guinea},\ and\
  \citenamefont {Castro~Neto}}]{Kotov12}%
  \BibitemOpen
  \bibfield  {author} {\bibinfo {author} {\bibfnamefont {Valeri~N.}\
  \bibnamefont {Kotov}}, \bibinfo {author} {\bibfnamefont {Bruno}\ \bibnamefont
  {Uchoa}}, \bibinfo {author} {\bibfnamefont {Vitor~M.}\ \bibnamefont
  {Pereira}}, \bibinfo {author} {\bibfnamefont {F.}~\bibnamefont {Guinea}}, \
  and\ \bibinfo {author} {\bibfnamefont {A.~H.}\ \bibnamefont {Castro~Neto}},\
  }\bibfield  {title} {\enquote {\bibinfo {title} {Electron-electron
  interactions in graphene: Current status and perspectives},}\ }\href
  {\doibase 10.1103/RevModPhys.84.1067} {\bibfield  {journal} {\bibinfo
  {journal} {Rev. Mod. Phys.}\ }\textbf {\bibinfo {volume} {84}},\ \bibinfo
  {pages} {1067--1125} (\bibinfo {year} {2012})}\BibitemShut {NoStop}%
\bibitem [{\citenamefont {Manzeli}\ \emph {et~al.}(2017)\citenamefont
  {Manzeli}, \citenamefont {Ovchinnikov}, \citenamefont {Pasquier},
  \citenamefont {Yazyev},\ and\ \citenamefont {Kis}}]{Manzeli17}%
  \BibitemOpen
  \bibfield  {author} {\bibinfo {author} {\bibfnamefont {S.}~\bibnamefont
  {Manzeli}}, \bibinfo {author} {\bibfnamefont {D.}~\bibnamefont
  {Ovchinnikov}}, \bibinfo {author} {\bibfnamefont {D.}~\bibnamefont
  {Pasquier}}, \bibinfo {author} {\bibfnamefont {O.V.}\ \bibnamefont {Yazyev}},
  \ and\ \bibinfo {author} {\bibfnamefont {A.}~\bibnamefont {Kis}},\ }\bibfield
   {title} {\enquote {\bibinfo {title} {{2D transition metal
  dichalcogenides}},}\ }\href {\doibase 10.1038/natrevmats.2017.33} {\bibfield
  {journal} {\bibinfo  {journal} {Nature Reviews Materials}\ }\textbf {\bibinfo
  {volume} {2}},\ \bibinfo {pages} {17033} (\bibinfo {year}
  {2017})}\BibitemShut {NoStop}%
\bibitem [{Note1()}]{Note1}%
  \BibitemOpen
  \bibinfo {note} {The reduced (Ising) order parameter symmetry of the Holstein
  case does support a finite $T$ transition in 2D.}\BibitemShut {Stop}%
\bibitem [{\citenamefont {Hirsch}(1987)}]{hirsch87}%
  \BibitemOpen
  \bibfield  {author} {\bibinfo {author} {\bibfnamefont {J.~E.}\ \bibnamefont
  {Hirsch}},\ }\bibfield  {title} {\enquote {\bibinfo {title} {{Simulations of
  the three-dimensional {H}ubbard model: Half-filled band sector}},}\ }\href
  {\doibase 10.1103/PhysRevB.35.1851} {\bibfield  {journal} {\bibinfo
  {journal} {Phys. Rev. B}\ }\textbf {\bibinfo {volume} {35}},\ \bibinfo
  {pages} {1851--1859} (\bibinfo {year} {1987})}\BibitemShut {NoStop}%
\bibitem [{\citenamefont {Scalettar}\ \emph {et~al.}(1989)\citenamefont
  {Scalettar}, \citenamefont {Scalapino}, \citenamefont {Sugar},\ and\
  \citenamefont {Toussaint}}]{scalettar89}%
  \BibitemOpen
  \bibfield  {author} {\bibinfo {author} {\bibfnamefont {R.~T.}\ \bibnamefont
  {Scalettar}}, \bibinfo {author} {\bibfnamefont {D.~J.}\ \bibnamefont
  {Scalapino}}, \bibinfo {author} {\bibfnamefont {R.~L.}\ \bibnamefont
  {Sugar}}, \ and\ \bibinfo {author} {\bibfnamefont {D.}~\bibnamefont
  {Toussaint}},\ }\bibfield  {title} {\enquote {\bibinfo {title} {{Phase
  diagram of the half-filled 3D Hubbard model}},}\ }\href {\doibase
  10.1103/PhysRevB.39.4711} {\bibfield  {journal} {\bibinfo  {journal} {Phys.
  Rev. B}\ }\textbf {\bibinfo {volume} {39}},\ \bibinfo {pages} {4711--4714}
  (\bibinfo {year} {1989})}\BibitemShut {NoStop}%
\bibitem [{\citenamefont {Staudt}\ \emph {et~al.}(2000)\citenamefont {Staudt},
  \citenamefont {Dzierzawa},\ and\ \citenamefont {Muramatsu}}]{Muramatsu00}%
  \BibitemOpen
  \bibfield  {author} {\bibinfo {author} {\bibfnamefont {R}~\bibnamefont
  {Staudt}}, \bibinfo {author} {\bibfnamefont {M.}~\bibnamefont {Dzierzawa}}, \
  and\ \bibinfo {author} {\bibfnamefont {A.}~\bibnamefont {Muramatsu}},\
  }\bibfield  {title} {\enquote {\bibinfo {title} {{Phase diagram of the
  three-dimensional Hubbard model at half filling}},}\ }\href {\doibase
  10.1007/s100510070120} {\bibfield  {journal} {\bibinfo  {journal} {Eur. Phys.
  J. B}\ }\textbf {\bibinfo {volume} {17}},\ \bibinfo {pages} {411--415}
  (\bibinfo {year} {2000})}\BibitemShut {NoStop}%
\bibitem [{\citenamefont {Fuchs}\ \emph {et~al.}(2011)\citenamefont {Fuchs},
  \citenamefont {Gull}, \citenamefont {Pollet}, \citenamefont {Burovski},
  \citenamefont {Kozik}, \citenamefont {Pruschke},\ and\ \citenamefont
  {Troyer}}]{fuchs11}%
  \BibitemOpen
  \bibfield  {author} {\bibinfo {author} {\bibfnamefont {Sebastian}\
  \bibnamefont {Fuchs}}, \bibinfo {author} {\bibfnamefont {Emanuel}\
  \bibnamefont {Gull}}, \bibinfo {author} {\bibfnamefont {Lode}\ \bibnamefont
  {Pollet}}, \bibinfo {author} {\bibfnamefont {Evgeni}\ \bibnamefont
  {Burovski}}, \bibinfo {author} {\bibfnamefont {Evgeny}\ \bibnamefont
  {Kozik}}, \bibinfo {author} {\bibfnamefont {Thomas}\ \bibnamefont
  {Pruschke}}, \ and\ \bibinfo {author} {\bibfnamefont {Matthias}\ \bibnamefont
  {Troyer}},\ }\bibfield  {title} {\enquote {\bibinfo {title} {{Thermodynamics
  of the 3{D} {H}ubbard model on approaching the {N}\'eel transition}},}\
  }\href {\doibase 10.1103/PhysRevLett.106.030401} {\bibfield  {journal}
  {\bibinfo  {journal} {Phys. Rev. Lett.}\ }\textbf {\bibinfo {volume} {106}},\
  \bibinfo {pages} {030401} (\bibinfo {year} {2011})}\BibitemShut {NoStop}%
\bibitem [{\citenamefont {Cohen-Stead}\ \emph {et~al.}(2020)\citenamefont
  {Cohen-Stead}, \citenamefont {Barros}, \citenamefont {Meng}, \citenamefont
  {Chen}, \citenamefont {Scalettar},\ and\ \citenamefont
  {Batrouni}}]{cohenstead20}%
  \BibitemOpen
  \bibfield  {author} {\bibinfo {author} {\bibfnamefont {B.}~\bibnamefont
  {Cohen-Stead}}, \bibinfo {author} {\bibfnamefont {Kipton}\ \bibnamefont
  {Barros}}, \bibinfo {author} {\bibfnamefont {ZY}~\bibnamefont {Meng}},
  \bibinfo {author} {\bibfnamefont {Chuang}\ \bibnamefont {Chen}}, \bibinfo
  {author} {\bibfnamefont {R.~T.}\ \bibnamefont {Scalettar}}, \ and\ \bibinfo
  {author} {\bibfnamefont {G.~G.}\ \bibnamefont {Batrouni}},\ }\bibfield
  {title} {\enquote {\bibinfo {title} {{Langevin simulations of the half-filled
  cubic Holstein model}},}\ }\href {\doibase 10.1103/PhysRevB.102.161108}
  {\bibfield  {journal} {\bibinfo  {journal} {Phys. Rev. B}\ }\textbf {\bibinfo
  {volume} {102}},\ \bibinfo {pages} {161108} (\bibinfo {year}
  {2020})}\BibitemShut {NoStop}%
\bibitem [{\citenamefont {Cohen-Stead}\ \emph {et~al.}()\citenamefont
  {Cohen-Stead}, \citenamefont {Barros}, \citenamefont {Scalettar},\ and\
  \citenamefont {Johnston}}]{cohenstead22}%
  \BibitemOpen
  \bibfield  {author} {\bibinfo {author} {\bibfnamefont {Benjamin}\
  \bibnamefont {Cohen-Stead}}, \bibinfo {author} {\bibfnamefont {Kipton}\
  \bibnamefont {Barros}}, \bibinfo {author} {\bibfnamefont {Richard}\
  \bibnamefont {Scalettar}}, \ and\ \bibinfo {author} {\bibfnamefont {Steven}\
  \bibnamefont {Johnston}},\ }\href@noop {} {\enquote {\bibinfo {title} {A
  hybrid {M}onte {C}arlo study of bond-stretching electron-phonon interactions
  and charge order in the bismuthate family of superconductors},}\ }\Eprint
  {http://arxiv.org/abs/cond-mat/2208.02339} {cond-mat/2208.02339} \BibitemShut
  {NoStop}%
\bibitem [{\citenamefont {McMahan}\ \emph {et~al.}(1998)\citenamefont
  {McMahan}, \citenamefont {Huscroft}, \citenamefont {Scalettar},\ and\
  \citenamefont {Pollock}}]{McMahan98}%
  \BibitemOpen
  \bibfield  {author} {\bibinfo {author} {\bibfnamefont {A.K.}\ \bibnamefont
  {McMahan}}, \bibinfo {author} {\bibfnamefont {C.}~\bibnamefont {Huscroft}},
  \bibinfo {author} {\bibfnamefont {R.T.}\ \bibnamefont {Scalettar}}, \ and\
  \bibinfo {author} {\bibfnamefont {E.L.}\ \bibnamefont {Pollock}},\ }\bibfield
   {title} {\enquote {\bibinfo {title} {Volume-collapse transitions in the rare
  earth metals},}\ }\href {\doibase 10.1023/A:1008698422183} {\bibfield
  {journal} {\bibinfo  {journal} {Journal of Computer-Aided Materials Design}\
  }\textbf {\bibinfo {volume} {5}},\ \bibinfo {pages} {131--162} (\bibinfo
  {year} {1998})}\BibitemShut {NoStop}%
\bibitem [{\citenamefont {Huscroft}\ \emph {et~al.}(1999)\citenamefont
  {Huscroft}, \citenamefont {McMahan},\ and\ \citenamefont
  {Scalettar}}]{Huscroft99}%
  \BibitemOpen
  \bibfield  {author} {\bibinfo {author} {\bibfnamefont {Carey}\ \bibnamefont
  {Huscroft}}, \bibinfo {author} {\bibfnamefont {A.~K.}\ \bibnamefont
  {McMahan}}, \ and\ \bibinfo {author} {\bibfnamefont {R.~T.}\ \bibnamefont
  {Scalettar}},\ }\bibfield  {title} {\enquote {\bibinfo {title} {Magnetic and
  thermodynamic properties of the three-dimensional periodic {A}nderson
  hamiltonian},}\ }\href {\doibase 10.1103/PhysRevLett.82.2342} {\bibfield
  {journal} {\bibinfo  {journal} {Phys. Rev. Lett.}\ }\textbf {\bibinfo
  {volume} {82}},\ \bibinfo {pages} {2342--2345} (\bibinfo {year}
  {1999})}\BibitemShut {NoStop}%
\bibitem [{\citenamefont {Paiva}\ \emph {et~al.}(2003)\citenamefont {Paiva},
  \citenamefont {Esirgen}, \citenamefont {Scalettar}, \citenamefont
  {Huscroft},\ and\ \citenamefont {McMahan}}]{Paiva03}%
  \BibitemOpen
  \bibfield  {author} {\bibinfo {author} {\bibfnamefont {Thereza}\ \bibnamefont
  {Paiva}}, \bibinfo {author} {\bibfnamefont {G\"okhan}\ \bibnamefont
  {Esirgen}}, \bibinfo {author} {\bibfnamefont {Richard~T.}\ \bibnamefont
  {Scalettar}}, \bibinfo {author} {\bibfnamefont {Carey}\ \bibnamefont
  {Huscroft}}, \ and\ \bibinfo {author} {\bibfnamefont {A.~K.}\ \bibnamefont
  {McMahan}},\ }\bibfield  {title} {\enquote {\bibinfo {title}
  {Doping-dependent study of the periodic {A}nderson model in three
  dimensions},}\ }\href {\doibase 10.1103/PhysRevB.68.195111} {\bibfield
  {journal} {\bibinfo  {journal} {Phys. Rev. B}\ }\textbf {\bibinfo {volume}
  {68}},\ \bibinfo {pages} {195111} (\bibinfo {year} {2003})}\BibitemShut
  {NoStop}%
\bibitem [{\citenamefont {Bernhard}\ and\ \citenamefont
  {Lacroix}(1999)}]{bernhard99}%
  \BibitemOpen
  \bibfield  {author} {\bibinfo {author} {\bibfnamefont {Ben~Hur}\ \bibnamefont
  {Bernhard}}\ and\ \bibinfo {author} {\bibfnamefont {Claudine}\ \bibnamefont
  {Lacroix}},\ }\bibfield  {title} {\enquote {\bibinfo {title} {{Thermodynamics
  of the Anderson lattice}},}\ }\href {\doibase 10.1103/PhysRevB.60.12149}
  {\bibfield  {journal} {\bibinfo  {journal} {Phys. Rev. B}\ }\textbf {\bibinfo
  {volume} {60}},\ \bibinfo {pages} {12149--12154} (\bibinfo {year}
  {1999})}\BibitemShut {NoStop}%
\bibitem [{\citenamefont {Held}\ \emph {et~al.}(2001)\citenamefont {Held},
  \citenamefont {McMahan},\ and\ \citenamefont {Scalettar}}]{held01}%
  \BibitemOpen
  \bibfield  {author} {\bibinfo {author} {\bibfnamefont {K.}~\bibnamefont
  {Held}}, \bibinfo {author} {\bibfnamefont {A.~K.}\ \bibnamefont {McMahan}}, \
  and\ \bibinfo {author} {\bibfnamefont {R.~T.}\ \bibnamefont {Scalettar}},\
  }\bibfield  {title} {\enquote {\bibinfo {title} {Cerium volume collapse:
  Results from the merger of dynamical mean-field theory and local density
  approximation},}\ }\href {\doibase 10.1103/PhysRevLett.87.276404} {\bibfield
  {journal} {\bibinfo  {journal} {Phys. Rev. Lett.}\ }\textbf {\bibinfo
  {volume} {87}},\ \bibinfo {pages} {276404} (\bibinfo {year}
  {2001})}\BibitemShut {NoStop}%
\bibitem [{\citenamefont {Blankenbecler}\ \emph {et~al.}(1981)\citenamefont
  {Blankenbecler}, \citenamefont {Scalapino},\ and\ \citenamefont
  {Sugar}}]{blankenbecler81}%
  \BibitemOpen
  \bibfield  {author} {\bibinfo {author} {\bibfnamefont {R.}~\bibnamefont
  {Blankenbecler}}, \bibinfo {author} {\bibfnamefont {D.~J.}\ \bibnamefont
  {Scalapino}}, \ and\ \bibinfo {author} {\bibfnamefont {R.~L.}\ \bibnamefont
  {Sugar}},\ }\bibfield  {title} {\enquote {\bibinfo {title} {{Monte Carlo
  calculations of coupled boson-fermion systems. I}},}\ }\href {\doibase
  10.1103/PhysRevD.24.2278} {\bibfield  {journal} {\bibinfo  {journal} {Phys.
  Rev. D}\ }\textbf {\bibinfo {volume} {24}},\ \bibinfo {pages} {2278--2286}
  (\bibinfo {year} {1981})}\BibitemShut {NoStop}%
\bibitem [{\citenamefont {Hirsch}(1985)}]{hirsch85}%
  \BibitemOpen
  \bibfield  {author} {\bibinfo {author} {\bibfnamefont {J.~E.}\ \bibnamefont
  {Hirsch}},\ }\bibfield  {title} {\enquote {\bibinfo {title} {{Two-dimensional
  Hubbard model: Numerical simulation study}},}\ }\href {\doibase
  10.1103/PhysRevB.31.4403} {\bibfield  {journal} {\bibinfo  {journal} {Phys.
  Rev. B}\ }\textbf {\bibinfo {volume} {31}},\ \bibinfo {pages} {4403--4419}
  (\bibinfo {year} {1985})}\BibitemShut {NoStop}%
\bibitem [{\citenamefont {White}\ \emph {et~al.}(1989)\citenamefont {White},
  \citenamefont {Scalapino}, \citenamefont {Sugar}, \citenamefont {Loh},
  \citenamefont {Gubernatis},\ and\ \citenamefont {Scalettar}}]{white89}%
  \BibitemOpen
  \bibfield  {author} {\bibinfo {author} {\bibfnamefont {S.~R.}\ \bibnamefont
  {White}}, \bibinfo {author} {\bibfnamefont {D.~J.}\ \bibnamefont
  {Scalapino}}, \bibinfo {author} {\bibfnamefont {R.~L.}\ \bibnamefont
  {Sugar}}, \bibinfo {author} {\bibfnamefont {E.~Y.}\ \bibnamefont {Loh}},
  \bibinfo {author} {\bibfnamefont {J.~E.}\ \bibnamefont {Gubernatis}}, \ and\
  \bibinfo {author} {\bibfnamefont {R.~T.}\ \bibnamefont {Scalettar}},\
  }\bibfield  {title} {\enquote {\bibinfo {title} {{Numerical study of the
  two-dimensional Hubbard model}},}\ }\href {\doibase 10.1103/PhysRevB.40.506}
  {\bibfield  {journal} {\bibinfo  {journal} {Phys. Rev. B}\ }\textbf {\bibinfo
  {volume} {40}},\ \bibinfo {pages} {506--516} (\bibinfo {year}
  {1989})}\BibitemShut {NoStop}%
\bibitem [{\citenamefont {\surname{dos Santos}}(2003)}]{dosSantos03b}%
  \BibitemOpen
  \bibfield  {author} {\bibinfo {author} {\bibfnamefont {Raimundo~R.}\
  \bibnamefont {\surname{dos Santos}}},\ }\bibfield  {title} {\enquote
  {\bibinfo {title} {{Introduction to quantum Monte Carlo simulations for
  fermionic systems}},}\ }\href {\doibase 10.1590/S0103-97332003000100003}
  {\bibfield  {journal} {\bibinfo  {journal} {Brazilian Journal of Physics}\
  }\textbf {\bibinfo {volume} {33}},\ \bibinfo {pages} {36 -- 54} (\bibinfo
  {year} {2003})}\BibitemShut {NoStop}%
\bibitem [{\citenamefont {Gubernatis}\ \emph {et~al.}(2016)\citenamefont
  {Gubernatis}, \citenamefont {Kawashima},\ and\ \citenamefont
  {Werner}}]{gubernatis16}%
  \BibitemOpen
  \bibfield  {author} {\bibinfo {author} {\bibfnamefont {J}~\bibnamefont
  {Gubernatis}}, \bibinfo {author} {\bibfnamefont {N}~\bibnamefont
  {Kawashima}}, \ and\ \bibinfo {author} {\bibfnamefont {P}~\bibnamefont
  {Werner}},\ }\href {\doibase 10.1143/PTPS.145.138} {\emph {\bibinfo {title}
  {Quantum Monte Carlo Methods: Algorithms for Lattice Models}}}\ (\bibinfo
  {publisher} {Cambridge University Press, Cambridge, England},\ \bibinfo
  {year} {2016})\BibitemShut {NoStop}%
\bibitem [{\citenamefont {Becca}\ and\ \citenamefont
  {Sorella}(2017)}]{becca17}%
  \BibitemOpen
  \bibfield  {author} {\bibinfo {author} {\bibfnamefont {Federico}\
  \bibnamefont {Becca}}\ and\ \bibinfo {author} {\bibfnamefont {Sandro}\
  \bibnamefont {Sorella}},\ }\href
  {https://www.cambridge.org/core/books/quantum-monte-carlo-approaches-for-correlated-systems/EB88C86BD9553A0738BDAE400D0B2900}
  {\emph {\bibinfo {title} {Quantum Monte Carlo approaches for correlated
  systems}}}\ (\bibinfo  {publisher} {Cambridge University Press, Cambridge,
  England},\ \bibinfo {year} {2017})\BibitemShut {NoStop}%
\bibitem [{\citenamefont {Loh}\ \emph {et~al.}(1990)\citenamefont {Loh},
  \citenamefont {Gubernatis}, \citenamefont {Scalettar}, \citenamefont {White},
  \citenamefont {Scalapino},\ and\ \citenamefont {Sugar}}]{Loh90}%
  \BibitemOpen
  \bibfield  {author} {\bibinfo {author} {\bibfnamefont {E.~Y.}\ \bibnamefont
  {Loh}}, \bibinfo {author} {\bibfnamefont {J.~E.}\ \bibnamefont {Gubernatis}},
  \bibinfo {author} {\bibfnamefont {R.~T.}\ \bibnamefont {Scalettar}}, \bibinfo
  {author} {\bibfnamefont {S.~R.}\ \bibnamefont {White}}, \bibinfo {author}
  {\bibfnamefont {D.~J.}\ \bibnamefont {Scalapino}}, \ and\ \bibinfo {author}
  {\bibfnamefont {R.~L.}\ \bibnamefont {Sugar}},\ }\bibfield  {title} {\enquote
  {\bibinfo {title} {{Sign problem in the numerical simulation of many-electron
  systems}},}\ }\href {\doibase 10.1103/PhysRevB.41.9301} {\bibfield  {journal}
  {\bibinfo  {journal} {Phys. Rev. B}\ }\textbf {\bibinfo {volume} {41}},\
  \bibinfo {pages} {9301--9307} (\bibinfo {year} {1990})}\BibitemShut {NoStop}%
\bibitem [{\citenamefont {Troyer}\ and\ \citenamefont
  {Wiese}(2005)}]{troyer05}%
  \BibitemOpen
  \bibfield  {author} {\bibinfo {author} {\bibfnamefont {Matthias}\
  \bibnamefont {Troyer}}\ and\ \bibinfo {author} {\bibfnamefont {Uwe-Jens}\
  \bibnamefont {Wiese}},\ }\bibfield  {title} {\enquote {\bibinfo {title}
  {Computational complexity and fundamental limitations to fermionic quantum
  {M}onte {C}arlo simulations},}\ }\href {\doibase
  10.1103/PhysRevLett.94.170201} {\bibfield  {journal} {\bibinfo  {journal}
  {Phys. Rev. Lett.}\ }\textbf {\bibinfo {volume} {94}},\ \bibinfo {pages}
  {170201} (\bibinfo {year} {2005})}\BibitemShut {NoStop}%
\bibitem [{\citenamefont {Mondaini}\ \emph {et~al.}(2022)\citenamefont
  {Mondaini}, \citenamefont {Tarat},\ and\ \citenamefont
  {Scalettar}}]{mondaini22}%
  \BibitemOpen
  \bibfield  {author} {\bibinfo {author} {\bibfnamefont {Rubem}\ \bibnamefont
  {Mondaini}}, \bibinfo {author} {\bibfnamefont {Sabyasachi}\ \bibnamefont
  {Tarat}}, \ and\ \bibinfo {author} {\bibfnamefont {Richard~T}\ \bibnamefont
  {Scalettar}},\ }\bibfield  {title} {\enquote {\bibinfo {title} {Quantum
  critical points and the sign problem},}\ }\href
  {https://www.science.org/doi/10.1126/science.abg9299} {\bibfield  {journal}
  {\bibinfo  {journal} {Science}\ }\textbf {\bibinfo {volume} {375}},\ \bibinfo
  {pages} {418--424} (\bibinfo {year} {2022})}\BibitemShut {NoStop}%
\bibitem [{\citenamefont {Kaul}(2015)}]{Kaul15}%
  \BibitemOpen
  \bibfield  {author} {\bibinfo {author} {\bibfnamefont {Ribhu~K.}\
  \bibnamefont {Kaul}},\ }\bibfield  {title} {\enquote {\bibinfo {title} {Spin
  nematics, valence-bond solids, and spin liquids in $\mathrm{SO}(n)$ quantum
  spin models on the triangular lattice},}\ }\href {\doibase
  10.1103/PhysRevLett.115.157202} {\bibfield  {journal} {\bibinfo  {journal}
  {Phys. Rev. Lett.}\ }\textbf {\bibinfo {volume} {115}},\ \bibinfo {pages}
  {157202} (\bibinfo {year} {2015})}\BibitemShut {NoStop}%
\bibitem [{\citenamefont {Darmawan}\ \emph {et~al.}(2018)\citenamefont
  {Darmawan}, \citenamefont {Nomura}, \citenamefont {Yamaji},\ and\
  \citenamefont {Imada}}]{Darmawan18}%
  \BibitemOpen
  \bibfield  {author} {\bibinfo {author} {\bibfnamefont {Andrew~S.}\
  \bibnamefont {Darmawan}}, \bibinfo {author} {\bibfnamefont {Yusuke}\
  \bibnamefont {Nomura}}, \bibinfo {author} {\bibfnamefont {Youhei}\
  \bibnamefont {Yamaji}}, \ and\ \bibinfo {author} {\bibfnamefont {Masatoshi}\
  \bibnamefont {Imada}},\ }\bibfield  {title} {\enquote {\bibinfo {title}
  {{Stripe and superconducting order competing in the Hubbard model on a square
  lattice studied by a combined variational Monte Carlo and tensor network
  method}},}\ }\href {\doibase 10.1103/PhysRevB.98.205132} {\bibfield
  {journal} {\bibinfo  {journal} {Phys. Rev. B}\ }\textbf {\bibinfo {volume}
  {98}},\ \bibinfo {pages} {205132} (\bibinfo {year} {2018})}\BibitemShut
  {NoStop}%
\bibitem [{\citenamefont {Paiva}\ \emph {et~al.}(2001)\citenamefont {Paiva},
  \citenamefont {Scalettar}, \citenamefont {Huscroft},\ and\ \citenamefont
  {McMahan}}]{paiva01}%
  \BibitemOpen
  \bibfield  {author} {\bibinfo {author} {\bibfnamefont {Thereza}\ \bibnamefont
  {Paiva}}, \bibinfo {author} {\bibfnamefont {R.~T.}\ \bibnamefont
  {Scalettar}}, \bibinfo {author} {\bibfnamefont {Carey}\ \bibnamefont
  {Huscroft}}, \ and\ \bibinfo {author} {\bibfnamefont {A.~K.}\ \bibnamefont
  {McMahan}},\ }\bibfield  {title} {\enquote {\bibinfo {title} {{Signatures of
  spin and charge energy scales in the local moment and specific heat of the
  half-filled two-dimensional Hubbard model}},}\ }\href {\doibase
  10.1103/PhysRevB.63.125116} {\bibfield  {journal} {\bibinfo  {journal} {Phys.
  Rev. B}\ }\textbf {\bibinfo {volume} {63}},\ \bibinfo {pages} {125116}
  (\bibinfo {year} {2001})}\BibitemShut {NoStop}%
\bibitem [{\citenamefont {Curro}(2009)}]{curro09}%
  \BibitemOpen
  \bibfield  {author} {\bibinfo {author} {\bibfnamefont {N.J.}\ \bibnamefont
  {Curro}},\ }\bibfield  {title} {\enquote {\bibinfo {title} {{Nuclear magnetic
  resonance in the heavy fermion superconductors}},}\ }\href {\doibase
  10.1088/0034-4885/72/2/026502} {\bibfield  {journal} {\bibinfo  {journal}
  {Reports on Progress in Physics}\ }\textbf {\bibinfo {volume} {72}},\
  \bibinfo {pages} {026502} (\bibinfo {year} {2009})}\BibitemShut {NoStop}%
\bibitem [{\citenamefont {Randeria}\ \emph {et~al.}(1992)\citenamefont
  {Randeria}, \citenamefont {Trivedi}, \citenamefont {Moreo},\ and\
  \citenamefont {Scalettar}}]{Randeria92}%
  \BibitemOpen
  \bibfield  {author} {\bibinfo {author} {\bibfnamefont {Mohit}\ \bibnamefont
  {Randeria}}, \bibinfo {author} {\bibfnamefont {Nandini}\ \bibnamefont
  {Trivedi}}, \bibinfo {author} {\bibfnamefont {Adriana}\ \bibnamefont
  {Moreo}}, \ and\ \bibinfo {author} {\bibfnamefont {Richard~T.}\ \bibnamefont
  {Scalettar}},\ }\bibfield  {title} {\enquote {\bibinfo {title} {{Pairing and
  spin gap in the normal state of short coherence length superconductors}},}\
  }\href {\doibase 10.1103/PhysRevLett.69.2001} {\bibfield  {journal} {\bibinfo
   {journal} {Phys. Rev. Lett.}\ }\textbf {\bibinfo {volume} {69}},\ \bibinfo
  {pages} {2001--2004} (\bibinfo {year} {1992})}\BibitemShut {NoStop}%
\bibitem [{\citenamefont {Duffy}\ and\ \citenamefont {Moreo}(1997)}]{Duffy97}%
  \BibitemOpen
  \bibfield  {author} {\bibinfo {author} {\bibfnamefont {Daniel}\ \bibnamefont
  {Duffy}}\ and\ \bibinfo {author} {\bibfnamefont {Adriana}\ \bibnamefont
  {Moreo}},\ }\bibfield  {title} {\enquote {\bibinfo {title} {{Specific heat of
  the two-dimensional Hubbard model}},}\ }\href {\doibase
  10.1103/PhysRevB.55.12918} {\bibfield  {journal} {\bibinfo  {journal} {Phys.
  Rev. B}\ }\textbf {\bibinfo {volume} {55}},\ \bibinfo {pages} {12918--12924}
  (\bibinfo {year} {1997})}\BibitemShut {NoStop}%
\bibitem [{\citenamefont {Veki\ifmmode~\acute{c}\else \'{c}\fi{}}\ \emph
  {et~al.}(1995)\citenamefont {Veki\ifmmode~\acute{c}\else \'{c}\fi{}},
  \citenamefont {Cannon}, \citenamefont {Scalapino}, \citenamefont
  {Scalettar},\ and\ \citenamefont {Sugar}}]{Vekic95}%
  \BibitemOpen
  \bibfield  {author} {\bibinfo {author} {\bibfnamefont {M.}~\bibnamefont
  {Veki\ifmmode~\acute{c}\else \'{c}\fi{}}}, \bibinfo {author} {\bibfnamefont
  {J.~W.}\ \bibnamefont {Cannon}}, \bibinfo {author} {\bibfnamefont {D.~J.}\
  \bibnamefont {Scalapino}}, \bibinfo {author} {\bibfnamefont {R.~T.}\
  \bibnamefont {Scalettar}}, \ and\ \bibinfo {author} {\bibfnamefont {R.~L.}\
  \bibnamefont {Sugar}},\ }\bibfield  {title} {\enquote {\bibinfo {title}
  {{Competition between Antiferromagnetic Order and Spin-Liquid Behavior in the
  Two-Dimensional Periodic Anderson Model at Half Filling}},}\ }\href {\doibase
  10.1103/PhysRevLett.74.2367} {\bibfield  {journal} {\bibinfo  {journal}
  {Phys. Rev. Lett.}\ }\textbf {\bibinfo {volume} {74}},\ \bibinfo {pages}
  {2367--2370} (\bibinfo {year} {1995})}\BibitemShut {NoStop}%
\bibitem [{\citenamefont {Hu}\ \emph {et~al.}(2017)\citenamefont {Hu},
  \citenamefont {Scalettar}, \citenamefont {Huang},\ and\ \citenamefont
  {Moritz}}]{hu17}%
  \BibitemOpen
  \bibfield  {author} {\bibinfo {author} {\bibfnamefont {Wenjian}\ \bibnamefont
  {Hu}}, \bibinfo {author} {\bibfnamefont {Richard~T.}\ \bibnamefont
  {Scalettar}}, \bibinfo {author} {\bibfnamefont {Edwin~W.}\ \bibnamefont
  {Huang}}, \ and\ \bibinfo {author} {\bibfnamefont {Brian}\ \bibnamefont
  {Moritz}},\ }\bibfield  {title} {\enquote {\bibinfo {title} {{Effects of an
  additional conduction band on the singlet-antiferromagnet competition in the
  periodic Anderson model}},}\ }\href {\doibase 10.1103/PhysRevB.95.235122}
  {\bibfield  {journal} {\bibinfo  {journal} {Phys. Rev. B}\ }\textbf {\bibinfo
  {volume} {95}},\ \bibinfo {pages} {235122} (\bibinfo {year}
  {2017})}\BibitemShut {NoStop}%
\bibitem [{\citenamefont {Sch\"afer}\ \emph {et~al.}(2019)\citenamefont
  {Sch\"afer}, \citenamefont {Katanin}, \citenamefont {Kitatani}, \citenamefont
  {Toschi},\ and\ \citenamefont {Held}}]{Schafer19}%
  \BibitemOpen
  \bibfield  {author} {\bibinfo {author} {\bibfnamefont {T.}~\bibnamefont
  {Sch\"afer}}, \bibinfo {author} {\bibfnamefont {A.~A.}\ \bibnamefont
  {Katanin}}, \bibinfo {author} {\bibfnamefont {M.}~\bibnamefont {Kitatani}},
  \bibinfo {author} {\bibfnamefont {A.}~\bibnamefont {Toschi}}, \ and\ \bibinfo
  {author} {\bibfnamefont {K.}~\bibnamefont {Held}},\ }\bibfield  {title}
  {\enquote {\bibinfo {title} {Quantum criticality in the two-dimensional
  periodic {A}nderson model},}\ }\href {\doibase
  10.1103/PhysRevLett.122.227201} {\bibfield  {journal} {\bibinfo  {journal}
  {Phys. Rev. Lett.}\ }\textbf {\bibinfo {volume} {122}},\ \bibinfo {pages}
  {227201} (\bibinfo {year} {2019})}\BibitemShut {NoStop}%
\bibitem [{\citenamefont {Mendes-Santos}\ \emph {et~al.}(2017)\citenamefont
  {Mendes-Santos}, \citenamefont {Costa}, \citenamefont {Batrouni},
  \citenamefont {Curro}, \citenamefont {dos Santos}, \citenamefont {Paiva},\
  and\ \citenamefont {Scalettar}}]{mendes-santos17}%
  \BibitemOpen
  \bibfield  {author} {\bibinfo {author} {\bibfnamefont {T.}~\bibnamefont
  {Mendes-Santos}}, \bibinfo {author} {\bibfnamefont {N.~C.}\ \bibnamefont
  {Costa}}, \bibinfo {author} {\bibfnamefont {G.}~\bibnamefont {Batrouni}},
  \bibinfo {author} {\bibfnamefont {N.}~\bibnamefont {Curro}}, \bibinfo
  {author} {\bibfnamefont {R.~R.}\ \bibnamefont {dos Santos}}, \bibinfo
  {author} {\bibfnamefont {T.}~\bibnamefont {Paiva}}, \ and\ \bibinfo {author}
  {\bibfnamefont {R.~T.}\ \bibnamefont {Scalettar}},\ }\bibfield  {title}
  {\enquote {\bibinfo {title} {{Impurities near an antiferromagnetic-singlet
  quantum critical point}},}\ }\href {\doibase 10.1103/PhysRevB.95.054419}
  {\bibfield  {journal} {\bibinfo  {journal} {Phys. Rev. B}\ }\textbf {\bibinfo
  {volume} {95}},\ \bibinfo {pages} {054419} (\bibinfo {year}
  {2017})}\BibitemShut {NoStop}%
\bibitem [{\citenamefont {Costa}\ \emph {et~al.}(2019)\citenamefont {Costa},
  \citenamefont {Mendes-Santos}, \citenamefont {Paiva}, \citenamefont {Curro},
  \citenamefont {dos Santos},\ and\ \citenamefont {Scalettar}}]{Costa19}%
  \BibitemOpen
  \bibfield  {author} {\bibinfo {author} {\bibfnamefont {N.~C.}\ \bibnamefont
  {Costa}}, \bibinfo {author} {\bibfnamefont {T.}~\bibnamefont
  {Mendes-Santos}}, \bibinfo {author} {\bibfnamefont {T.}~\bibnamefont
  {Paiva}}, \bibinfo {author} {\bibfnamefont {N.~J.}\ \bibnamefont {Curro}},
  \bibinfo {author} {\bibfnamefont {R.~R.}\ \bibnamefont {dos Santos}}, \ and\
  \bibinfo {author} {\bibfnamefont {R.~T.}\ \bibnamefont {Scalettar}},\
  }\bibfield  {title} {\enquote {\bibinfo {title} {Coherence temperature in the
  diluted periodic {A}nderson model},}\ }\href {\doibase
  10.1103/PhysRevB.99.195116} {\bibfield  {journal} {\bibinfo  {journal} {Phys.
  Rev. B}\ }\textbf {\bibinfo {volume} {99}},\ \bibinfo {pages} {195116}
  (\bibinfo {year} {2019})}\BibitemShut {NoStop}%
\bibitem [{\citenamefont {Slater}(1951)}]{slater51}%
  \BibitemOpen
  \bibfield  {author} {\bibinfo {author} {\bibfnamefont {J.C.}\ \bibnamefont
  {Slater}},\ }\bibfield  {title} {\enquote {\bibinfo {title} {Magnetic effects
  and the {H}artree-{F}ock equation},}\ }\href
  {https://journals.aps.org/pr/abstract/10.1103/PhysRev.82.538} {\bibfield
  {journal} {\bibinfo  {journal} {Physical Review}\ }\textbf {\bibinfo {volume}
  {82}},\ \bibinfo {pages} {538} (\bibinfo {year} {1951})}\BibitemShut
  {NoStop}%
\bibitem [{\citenamefont {Nagaosa}(1999)}]{nagaosa99}%
  \BibitemOpen
  \bibfield  {author} {\bibinfo {author} {\bibfnamefont {Naoto}\ \bibnamefont
  {Nagaosa}},\ }\href {https://link.springer.com/book/9783540659815} {\emph
  {\bibinfo {title} {Quantum field theory in strongly correlated electronic
  systems}}}\ (\bibinfo  {publisher} {Springer Science \& Business Media},\
  \bibinfo {year} {1999})\BibitemShut {NoStop}%
\bibitem [{\citenamefont {Kozik}\ \emph {et~al.}(2013)\citenamefont {Kozik},
  \citenamefont {Burovski}, \citenamefont {Scarola},\ and\ \citenamefont
  {Troyer}}]{kozik13}%
  \BibitemOpen
  \bibfield  {author} {\bibinfo {author} {\bibfnamefont {E.}~\bibnamefont
  {Kozik}}, \bibinfo {author} {\bibfnamefont {E.}~\bibnamefont {Burovski}},
  \bibinfo {author} {\bibfnamefont {V.~W.}\ \bibnamefont {Scarola}}, \ and\
  \bibinfo {author} {\bibfnamefont {M.}~\bibnamefont {Troyer}},\ }\bibfield
  {title} {\enquote {\bibinfo {title} {N\'eel temperature and thermodynamics of
  the half-filled three-dimensional {H}ubbard model by diagrammatic determinant
  {M}onte {C}arlo},}\ }\href {\doibase 10.1103/PhysRevB.87.205102} {\bibfield
  {journal} {\bibinfo  {journal} {Phys. Rev. B}\ }\textbf {\bibinfo {volume}
  {87}},\ \bibinfo {pages} {205102} (\bibinfo {year} {2013})}\BibitemShut
  {NoStop}%
\bibitem [{\citenamefont {Hirschmeier}\ \emph {et~al.}(2015)\citenamefont
  {Hirschmeier}, \citenamefont {Hafermann}, \citenamefont {Gull}, \citenamefont
  {Lichtenstein},\ and\ \citenamefont {Antipov}}]{Hirschmeier15}%
  \BibitemOpen
  \bibfield  {author} {\bibinfo {author} {\bibfnamefont {Daniel}\ \bibnamefont
  {Hirschmeier}}, \bibinfo {author} {\bibfnamefont {Hartmut}\ \bibnamefont
  {Hafermann}}, \bibinfo {author} {\bibfnamefont {Emanuel}\ \bibnamefont
  {Gull}}, \bibinfo {author} {\bibfnamefont {Alexander~I.}\ \bibnamefont
  {Lichtenstein}}, \ and\ \bibinfo {author} {\bibfnamefont {Andrey~E.}\
  \bibnamefont {Antipov}},\ }\bibfield  {title} {\enquote {\bibinfo {title}
  {{Mechanisms of finite-temperature magnetism in the three-dimensional Hubbard
  model}},}\ }\href {\doibase 10.1103/PhysRevB.92.144409} {\bibfield  {journal}
  {\bibinfo  {journal} {Phys. Rev. B}\ }\textbf {\bibinfo {volume} {92}},\
  \bibinfo {pages} {144409} (\bibinfo {year} {2015})}\BibitemShut {NoStop}%
\bibitem [{\citenamefont {Khatami}(2016)}]{Khatami16}%
  \BibitemOpen
  \bibfield  {author} {\bibinfo {author} {\bibfnamefont {Ehsan}\ \bibnamefont
  {Khatami}},\ }\bibfield  {title} {\enquote {\bibinfo {title}
  {{Three-dimensional Hubbard model in the thermodynamic limit}},}\ }\href
  {\doibase 10.1103/PhysRevB.94.125114} {\bibfield  {journal} {\bibinfo
  {journal} {Phys. Rev. B}\ }\textbf {\bibinfo {volume} {94}},\ \bibinfo
  {pages} {125114} (\bibinfo {year} {2016})}\BibitemShut {NoStop}%
\bibitem [{Note2()}]{Note2}%
  \BibitemOpen
  \bibinfo {note} {Interestingly, the N\'eel temperatures obtained by
  DMFT\protect \,\cite {Held00,Schafer19} are compatible with our QMC
  ones.}\BibitemShut {Stop}%
\bibitem [{\citenamefont {Held}\ \emph {et~al.}(2000)\citenamefont {Held},
  \citenamefont {Huscroft}, \citenamefont {Scalettar},\ and\ \citenamefont
  {McMahan}}]{Held00}%
  \BibitemOpen
  \bibfield  {author} {\bibinfo {author} {\bibfnamefont {K.}~\bibnamefont
  {Held}}, \bibinfo {author} {\bibfnamefont {C.}~\bibnamefont {Huscroft}},
  \bibinfo {author} {\bibfnamefont {R.~T.}\ \bibnamefont {Scalettar}}, \ and\
  \bibinfo {author} {\bibfnamefont {A.~K.}\ \bibnamefont {McMahan}},\
  }\bibfield  {title} {\enquote {\bibinfo {title} {Similarities between the
  {H}ubbard and periodic {A}nderson models at finite temperatures},}\ }\href
  {\doibase 10.1103/PhysRevLett.85.373} {\bibfield  {journal} {\bibinfo
  {journal} {Phys. Rev. Lett.}\ }\textbf {\bibinfo {volume} {85}},\ \bibinfo
  {pages} {373--376} (\bibinfo {year} {2000})}\BibitemShut {NoStop}%
\bibitem [{\citenamefont {Kim}\ \emph {et~al.}(2020)\citenamefont {Kim},
  \citenamefont {Simkovic},\ and\ \citenamefont {Kozik}}]{Kim20}%
  \BibitemOpen
  \bibfield  {author} {\bibinfo {author} {\bibfnamefont {Aaram~J.}\
  \bibnamefont {Kim}}, \bibinfo {author} {\bibfnamefont {Fedor}\ \bibnamefont
  {Simkovic}}, \ and\ \bibinfo {author} {\bibfnamefont {Evgeny}\ \bibnamefont
  {Kozik}},\ }\bibfield  {title} {\enquote {\bibinfo {title} {Spin and charge
  correlations across the metal-to-insulator crossover in the half-filled 2{D}
  {H}ubbard model},}\ }\href {\doibase 10.1103/PhysRevLett.124.117602}
  {\bibfield  {journal} {\bibinfo  {journal} {Phys. Rev. Lett.}\ }\textbf
  {\bibinfo {volume} {124}},\ \bibinfo {pages} {117602} (\bibinfo {year}
  {2020})}\BibitemShut {NoStop}%
\bibitem [{\citenamefont {Trivedi}\ \emph {et~al.}(1996)\citenamefont
  {Trivedi}, \citenamefont {Scalettar},\ and\ \citenamefont
  {Randeria}}]{Trivedi96}%
  \BibitemOpen
  \bibfield  {author} {\bibinfo {author} {\bibfnamefont {Nandini}\ \bibnamefont
  {Trivedi}}, \bibinfo {author} {\bibfnamefont {Richard~T.}\ \bibnamefont
  {Scalettar}}, \ and\ \bibinfo {author} {\bibfnamefont {Mohit}\ \bibnamefont
  {Randeria}},\ }\bibfield  {title} {\enquote {\bibinfo {title}
  {Superconductor-insulator transition in a disordered electronic system},}\
  }\href {\doibase 10.1103/PhysRevB.54.R3756} {\bibfield  {journal} {\bibinfo
  {journal} {Phys. Rev. B}\ }\textbf {\bibinfo {volume} {54}},\ \bibinfo
  {pages} {R3756--R3759} (\bibinfo {year} {1996})}\BibitemShut {NoStop}%
\bibitem [{\citenamefont {Denteneer}\ \emph {et~al.}(1999)\citenamefont
  {Denteneer}, \citenamefont {Scalettar},\ and\ \citenamefont
  {Trivedi}}]{Denteneer99}%
  \BibitemOpen
  \bibfield  {author} {\bibinfo {author} {\bibfnamefont {P.~J.~H.}\
  \bibnamefont {Denteneer}}, \bibinfo {author} {\bibfnamefont {R.~T.}\
  \bibnamefont {Scalettar}}, \ and\ \bibinfo {author} {\bibfnamefont
  {N.}~\bibnamefont {Trivedi}},\ }\bibfield  {title} {\enquote {\bibinfo
  {title} {Conducting phase in the two-dimensional disordered {H}ubbard
  model},}\ }\href {\doibase 10.1103/PhysRevLett.83.4610} {\bibfield  {journal}
  {\bibinfo  {journal} {Phys. Rev. Lett.}\ }\textbf {\bibinfo {volume} {83}},\
  \bibinfo {pages} {4610--4613} (\bibinfo {year} {1999})}\BibitemShut {NoStop}%
\bibitem [{\citenamefont {Mondaini}\ \emph {et~al.}(2012)\citenamefont
  {Mondaini}, \citenamefont {Bouadim}, \citenamefont {Paiva},\ and\
  \citenamefont {dos Santos}}]{Mondaini12}%
  \BibitemOpen
  \bibfield  {author} {\bibinfo {author} {\bibfnamefont {Rubem}\ \bibnamefont
  {Mondaini}}, \bibinfo {author} {\bibfnamefont {K.}~\bibnamefont {Bouadim}},
  \bibinfo {author} {\bibfnamefont {Thereza}\ \bibnamefont {Paiva}}, \ and\
  \bibinfo {author} {\bibfnamefont {Raimundo~R.}\ \bibnamefont {dos Santos}},\
  }\bibfield  {title} {\enquote {\bibinfo {title} {Finite-size effects in
  transport data from quantum {M}onte {C}arlo simulations},}\ }\href {\doibase
  10.1103/PhysRevB.85.125127} {\bibfield  {journal} {\bibinfo  {journal} {Phys.
  Rev. B}\ }\textbf {\bibinfo {volume} {85}},\ \bibinfo {pages} {125127}
  (\bibinfo {year} {2012})}\BibitemShut {NoStop}%
\bibitem [{Note3()}]{Note3}%
  \BibitemOpen
  \bibinfo {note} {We expect that, for larger lattice sizes, the low-$T$ peak
  would be enhanced for $V < V_c$, and suppressed for $V > V_c$, emphasizing
  the transition.}\BibitemShut {Stop}%
\bibitem [{\citenamefont {Wu}\ and\ \citenamefont {Tremblay}(2015)}]{Wu15}%
  \BibitemOpen
  \bibfield  {author} {\bibinfo {author} {\bibfnamefont {Wei}\ \bibnamefont
  {Wu}}\ and\ \bibinfo {author} {\bibfnamefont {A.-M.-S.}\ \bibnamefont
  {Tremblay}},\ }\bibfield  {title} {\enquote {\bibinfo {title} {{$d$-wave
  superconductivity in the frustrated two-dimensional periodic Anderson
  model}},}\ }\href {\doibase 10.1103/PhysRevX.5.011019} {\bibfield  {journal}
  {\bibinfo  {journal} {Phys. Rev. X}\ }\textbf {\bibinfo {volume} {5}},\
  \bibinfo {pages} {011019} (\bibinfo {year} {2015})}\BibitemShut {NoStop}%
\bibitem [{\citenamefont {Assaad}(2002)}]{Assaad02}%
  \BibitemOpen
  \bibfield  {author} {\bibinfo {author} {\bibfnamefont {F.~F.}\ \bibnamefont
  {Assaad}},\ }\bibfield  {title} {\enquote {\bibinfo {title} {{Depleted Kondo
  lattices: Quantum Monte Carlo and mean-field calculations}},}\ }\href
  {\doibase 10.1103/PhysRevB.65.115104} {\bibfield  {journal} {\bibinfo
  {journal} {Phys. Rev. B}\ }\textbf {\bibinfo {volume} {65}},\ \bibinfo
  {pages} {115104} (\bibinfo {year} {2002})}\BibitemShut {NoStop}%
\bibitem [{\citenamefont {Costa}\ \emph {et~al.}(2018)\citenamefont {Costa},
  \citenamefont {Ara\'ujo}, \citenamefont {Lima}, \citenamefont {Paiva},
  \citenamefont {dos Santos},\ and\ \citenamefont {Scalettar}}]{costa18a}%
  \BibitemOpen
  \bibfield  {author} {\bibinfo {author} {\bibfnamefont {N.~C.}\ \bibnamefont
  {Costa}}, \bibinfo {author} {\bibfnamefont {M.~V.}\ \bibnamefont {Ara\'ujo}},
  \bibinfo {author} {\bibfnamefont {J.~P.}\ \bibnamefont {Lima}}, \bibinfo
  {author} {\bibfnamefont {T.}~\bibnamefont {Paiva}}, \bibinfo {author}
  {\bibfnamefont {R.~R.}\ \bibnamefont {dos Santos}}, \ and\ \bibinfo {author}
  {\bibfnamefont {R.~T.}\ \bibnamefont {Scalettar}},\ }\bibfield  {title}
  {\enquote {\bibinfo {title} {{Compressible ferrimagnetism in the depleted
  periodic Anderson model}},}\ }\href {\doibase 10.1103/PhysRevB.97.085123}
  {\bibfield  {journal} {\bibinfo  {journal} {Phys. Rev. B}\ }\textbf {\bibinfo
  {volume} {97}},\ \bibinfo {pages} {085123} (\bibinfo {year}
  {2018})}\BibitemShut {NoStop}%
\bibitem [{\citenamefont {Jiang}(2020)}]{Jiang20}%
  \BibitemOpen
  \bibfield  {author} {\bibinfo {author} {\bibfnamefont {Mi}~\bibnamefont
  {Jiang}},\ }\bibfield  {title} {\enquote {\bibinfo {title} {Enhanced tendency
  towards $d$-wave pairing and antiferromagnetism in a doped staggered periodic
  {A}nderson model},}\ }\href {\doibase 10.1103/PhysRevB.102.085119} {\bibfield
   {journal} {\bibinfo  {journal} {Phys. Rev. B}\ }\textbf {\bibinfo {volume}
  {102}},\ \bibinfo {pages} {085119} (\bibinfo {year} {2020})}\BibitemShut
  {NoStop}%
\end{thebibliography}%


\end{document}